\documentclass[11pt,onecolumn]{IEEEtran}
\usepackage{mathpple}
\usepackage{times}
\usepackage{amsmath}  
\usepackage{amssymb}  
\usepackage{mathrsfs} 
\usepackage{theorem}  
\usepackage{cite}     
\usepackage{comment}  
\usepackage{upref}
\usepackage{amsfonts}
\usepackage{verbatim}
\usepackage[dvipsnames,usenames]{color}
\usepackage{enumerate}
\usepackage{graphicx}
\usepackage{subfigure}
\usepackage{latexsym}

\usepackage{psfrag}

\usepackage{color}
\usepackage{times,color}
\usepackage[usenames,dvipsnames,svgnames,table]{xcolor}

\usepackage{pgf}
\usepackage{tikz}
\usetikzlibrary{arrows, automata, positioning, calc}

\usepackage{cancel} 

\usepackage{algorithm,algpseudocode}

\makeatletter
\newcommand{\removelatexerror}{\let\@latex@error\@gobble}
\makeatletter
\newcommand{\proofpart}[2]{%
	\par
	\addvspace{\medskipamount}%
	\noindent\emph{Part #1: #2}\par\nobreak
	\addvspace{\smallskipamount}%
	\@afterheading
}
\makeatother

\newcommand{\code}{\ttfamily\bfseries}

\newcommand{\be}[1]{\begin{equation}\label{#1}}
\newcommand{\ee}{\end{equation}}

\newcommand{\bc}{\begin{center}}
\newcommand{\ec}{\end{center}}

\newcommand{\qed}{\hfill$\Box$\\[1ex]}

\newcommand{\cA}{{\cal A}}
\newcommand{\cB}{{\cal B}}

\newcommand{\cG}{{\cal G}}

\newcommand{\cQ}{{\cal Q}}
\newcommand{\cR}{{\cal R}}

\newcommand{\cV}{{\cal V}}
\newcommand{\cW}{{\cal W}}
\newcommand{\cX}{{\cal X}}
\newcommand{\cY}{{\cal Y}}

\newcommand{\bzero}{{\mathbf 0}}

\renewcommand{\le}{\leqslant}
\renewcommand{\leq}{\leqslant}
\renewcommand{\ge}{\geqslant}
\renewcommand{\geq}{\geqslant}

\newcommand{\deff}{\mbox{$\stackrel{\rm def}{=}$}}

\newcommand{\F}{\mathbb{F}}

\DeclareMathOperator{\supp}{supp}

\newcommand{\Cref}[1]{Co\-rol\-la\-ry\,\ref{#1}}

\theoremstyle{plain} \theorembodyfont{\normalfont\slshape}

\newtheorem{thm}{Theorem$\!$}
\newenvironment{theorem}{\begin{thm}\hspace*{-1ex}{\bf.}}{\end{thm}}

\newtheorem{prop}[thm]{Proposition$\!$}
\newenvironment{proposition}{\begin{prop}\hspace*{-1ex}{\bf.}}{\end{prop}}

\newtheorem{lem}[thm]{Lemma$\!$}
\newenvironment{lemma}{\begin{lem}\hspace*{-1ex}{\bf.}}{\end{lem}}

\newtheorem{cor}[thm]{Corollary$\!$}
\newenvironment{corollary}{\begin{cor}\hspace*{-1ex}{\bf.}}{\end{cor}}

\newtheorem{const}[thm]{Construction$\!$}

\newtheorem{conj}[thm]{Conjecture$\!$}
\newenvironment{conjecture}{\begin{conj}\hspace*{-1ex}{\bf.}}{\end{conj}}

\newtheorem{defi}[thm]{Definition$\!$}
\newenvironment{definition}{\begin{defi}\hspace*{-1ex}{\bf.}}{\end{defi}}

\theorembodyfont{\normalfont}
\newtheorem{exam}{Example$\!$}
\newenvironment{example}{\begin{exam}\hspace*{-1ex}{\bf .}}{\qed\end{exam}}

\newtheorem{remrk}{Remark$\!$}
\newenvironment{remark}{\begin{remrk}\hspace*{-1ex}{\bf .}}{\end{remrk}}

\definecolor{Codecolor}{named}{White}  

\newcommand{\Copen}{\mbox{\{\kern-5.50pt\{}}
\newcommand{\Cclose}{\mbox{\}\kern-5.50pt\}}}
\newcommand{\Cslash}{\mbox{$\backslash\kern-6.02pt\backslash$}}

\makeatletter

\DeclareRobustCommand{\sbinom}{\genfrac[]\z@{}}

\makeatother

\newcommand{\G}[2]{\sbinom{{#1}\kern-.05pt}{{#2}\kern-.05pt}}

\usepackage{soul}
\setul{}{2pt}
\colorlet{C1}{cyan!50}

\begin{document}

\title{\textbf{Private Proximity Retrieval Codes}}

\author{
\textbf{Yiwei Zhang}, 
\textbf{Eitan Yaakobi}, 
and \textbf{Tuvi Etzion}
\thanks{The authors are with the Department of Computer Science, Technion -- Israel Institute of Technology, Haifa 3200003, Israel, (e-mail: \{ywzhang,yaakobi,etzion\}@cs.technion.ac.il).}
\thanks{Part of the results in this paper were presented at the IEEE International Symposium on Information Theory (ISIT 2019), Paris, France, July 2019 (reference~\cite{EGKYZ}).}%
\thanks{Eitan Yaakobi and Yiwei Zhang were supported in part by the ISF grant 1817/18;
Tuvi Etzion and Yiwei Zhang were supported in part by the BSF-NSF grant 2016692;
Yiwei Zhang was also supported in part by a Technion Fellowship. This research was partially supported by the Technion Hiroshi Fujiwara cyber security research center and the Israel cyber directorate.}
}

\maketitle

\begin{abstract}
A \emph{private proximity retrieval} (\emph{PPR}) scheme is a protocol which allows a user to retrieve the identities of all records in a database
that are within some distance $r$ from the user's record $x$. The user's \emph{privacy} at each server is given by the fraction
of the record $x$ that is kept private. In this paper,
this research is initiated and protocols that offer trade-offs between privacy and computational complexity and storage are studied.
In particular, we assume that each server stores a copy of the database and study the required minimum number of servers by our protocol which provides a given privacy level.
Each server gets a query in the protocol and the set of queries forms a code. We study the family of codes generated by the set of
queries and in particular the minimum number of codewords in such a code which is the minimum number of servers
required for the protocol. These codes are closely related to a family of codes known as \emph{covering designs}.
We introduce several lower bounds on the sizes of such codes as well as several constructions.
This work focuses on the case when the records are binary vectors together with the Hamming distance.
Other metrics such as the Johnson metric are also investigated.
\end{abstract}



\section{Introduction}
The growing amount of available information, which mostly resides in the web and on the cloud, has made \emph{information retrieval} (\emph{IR}) one of the more important computing tasks. In fact, web search has become the standard source of information search and in general, IR refers to information access in order to obtain data, in any form, from any available information resources. However, this form of communication also poses a risk to the user privacy, since the servers can monitor the user's requests in order to deduce important information on the user and his interests. Therefore, an important aspect of IR is hiding the information the user is searching for.

\subsection{Private Information Retrieval, Private Computation, and Proximity Searching}
 \emph{Private information retrieval} (PIR) is one of the well-known studied problems that provide privacy to user's requests. This problem was introduced by Chor, Goldreich, Kushilevitz, and Sudan in \cite{CKGS98}.  PIR protocols make it possible to retrieve a data item from a database without disclosing any information about the identity of the item being retrieved. 
This problem has attracted considerable attention since its inception, see\cite{BIK05,BIKR02,BIM00,CHY14,CHY15,DG14,G04,WY05,Y10,Y08}. The classic PIR model of~\cite{CKGS98} views the database as a collection of bits (or records) and assumes that the user wishes to retrieve the $i$th record without revealing any information about the index $i$. A naive solution is to download the entire database, and in fact this solution cannot be improved upon if the database is stored on a single server~\cite{CKGS98}.
In order to achieve sublinear communication complexity it was proposed in~\cite{CKGS98} to \emph{replicate the database} on several servers that do not communicate with each other. One of the disadvantages of this PIR model is that every record consists of a single bit, while in practice a record is much longer. As a consequence of this model's drawback, this problem has received recently significant attention from an information-theoretic perspective, wherein the database consists of large records (the number of bits in each file is much larger than the number of files) and the goal is to minimize the number of bits that are downloaded from the servers~\cite{CHY14,CHY15}. 
This reformulation of the problem introduces the \emph{rate} of a PIR scheme to be the ratio between the size of the file and the total number of downloaded bits from all servers~\cite{SJ17A}. Since then, extensions of this model for several more setups have been rigorously studied; see e.g.~\cite{SJ17A,SJ18A,BU18A,BU18B,TGR18,FGHK17,KRA16,WS16} and references therein.

PIR protocols have been rigorously studied mostly for the very basic IR problem which assumes that the user knows which data records are stored in the database (but not their content) and simply asks for the content of one of them.  However, the user may be interested in other forms of information from the database rather than just asking for a particular record. The recently-introduced problem of \emph{private computation} is a generalization of PIR which allows a user to compute an arbitrary function of a database, without revealing the identity of the function to the database.  This problem has been studied for linear functions in \cite{SJ17B,MAMA17,OLRK18} and for polynomial functions in \cite{Dave18,RK18}.  Many open questions about private computation remain, especially for non-linear functions, one of which is the topic of this paper.

Another important IR problem is that of \emph{proximity searching}. One example of proximity searching is the \emph{${K\textmd{-nearest}}$ neighbor search} (\emph{$K$-NNS}), where the goal is to find the $K$ elements from the database that minimize the distance to a given query~\cite{CFN08,MY16}. Proximity searching has several applications, among them are classification, searching for similar objects in multimedia databases, searching for similar documents in information retrieval, similar biological sequences in computational biology, and more. While proximity searching has been well-studied in the literature, to the best of our knowledge, there are no existing solutions which offer both proximity searching and privacy simultaneously. The assumption in proximity searching algorithms is that there is only a single server which stores the database. However, modern data storage systems are stored across several servers in a distributed manner, and thus we can take advantage of this setup in order to provide privacy for proximity retrieval.

\subsection{Private Proximity Retrieval}
In our setup of the problem we assume the user has a record and is only interested in knowing the identities, i.e., the indices, of the records in the database that are close to his record. The measure of closeness is determined according to some distance measure. This setup can fit for example to the case when the records stored in the database are attributes of different users. Given the user's attributes, he is only interested to know the users which have similar attributes to his according to some distance measure between the attributes. For example, a record may be a user's location and the database may consist of the locations of agents.  In this context, the user seeks to determine the identity of the agents nearest to him, without necessarily knowing their exact locations, while minimizing the amount of information that is exposed to the servers about his location. This example is related to the \emph{private proximity testing} problem in which two users seek to determine whether they are close to each other, without revealing any information about each other's location~\cite{NTLHB11,NPS12,PKB15}. Yet another example assumes that each record is a file (song, movie, DNA sequence etc.), and the user is interested in determining the records which are similar to his.


Assume an $M$-file database $\mathcal{X}=\{ x_1,\ldots,x_M \}\subseteq \cV$ is stored on $N$ non-colluding servers and the user has a record $x\in \cV$. Let $B(x,r)$ be the ball of radius $r$ centered at $x$ according to some distance measure $d$ {on~$\cV$}. The user seeks to know the identity of every record in $B(x,r)\cap\mathcal{X}$, i.e., the index of every record in the database similar to his, which is given by the set $I(x,r) \triangleq \{m\in \{1,\ldots,M\} |\ x_m\in B(x,r)\}$. As opposed to the classical PIR problem, our solutions do not provide full privacy. Thus, it is required that this computation is invoked while minimizing the amount of information the servers learn about $x$.

A \emph{private proximity retrieval} (\emph{PPR}) scheme consists of three steps. First, the user generates, with some degree of randomness, $N$ queries which are sent to the $N$ servers. Based upon these queries and the database $\cX$, each server responds back with an answer to the user, and finally the user calculates the set $I(x,r)$. The user's privacy at each server is defined to be the fraction of bits of $x$ which are kept private from the server. The schemes we study in this paper have a common structure. The query sent to the $n$th server is given by $q_n=(y_n,\rho_n)$, where $y_n$ is an element of $\cV$ and $\rho_n$ is a distance parameter, and the server's answer is the set of indices given by $A_n =I(y_n,\rho_n)$. Finally, the user computes the set $I(x,r) = \bigcap_{n\in[N]} A_n = \bigcap_{n\in[N]}I(y_n,\rho_n)$.  The main goal of our work is studying the trade-offs between the search radius $r$, the privacy at each server, and the number of servers.  In general, one can achieve more privacy at the expense of a smaller search radius or the use of more servers.
Finally, it is worth noting that private proximity retrieval can be viewed as another instance of the general private computation problem, in which $I(x,r)$ is the function to be computed on $\mathcal{X}$.  Here the user wishes to hide the identity of this function amongst all functions of the form $I(y,r)$ for any $y\in\mathcal{V}$.

While the study of PPR schemes can be investigated for arbitrary space $\cV$ and distance measure $d$ on $\cV$, the primary focus of this paper is on the case $\cV=\F_2^L$, where $d$ is the Hamming distance on $\F_2^L$. Then, the tradeoff between the number of servers and the privacy can be translated to the problem of finding the smallest subset $C$ of $\cW_s \triangleq \{ y\in \F_2^L \ | \ w(y) = s \}$ such that
$$B(\bzero,r) = \bigcap_{c\in C}B(c,r+s),$$
where $s$ is a design parameter of the PPR scheme which determines its privacy and $w(y)$ is the Hamming weight of $y$. Such a subset $C$ will be called an $(L,s,r)$ \emph{Private Proximity Retrieval Intersection Covering (PPRIC) code}, and the minimum size of $C$ will be denoted by $N(L,s,r)$. Our primary interest in the paper is to determine exactly the value of $N(L,s,r)$ for a large range of the parameters $L,s,r$ by providing tight lower and upper bounds on this value. An important connection in studying the value of $N(L,s,r)$ is derived by showing that PPRIC codes are closely related to a well-studied problem in design theory, called \emph{covering design}. Namely, an $(n,k,t)$ covering design, $n \geq k \geq t > 0$, is a collection of $k$-subsets (called \emph{blocks}) of an $n$-set
such that each $t$-subset of the $n$-set is contained in at least one block of the collection. We will show that PPRIC codes are also covering codes which have some additional constraints. Thus, the existence of a PPRIC code implies a corresponding covering design, but the opposite direction does not always hold. Therefore, all lower bounds on covering designs hold also for PPRIC codes and yet in many cases the constructions of covering designs are still valid for PPRIC codes. For the remaining parameters we show how to close on the gap by the results that can be derived by covering designs and present tighter lower bounds and specific code constructions.

\subsection{Paper Organization}

The rest of this paper is organized as follows. In Section~\ref{sec:protocol}, the PPR problem is formally defined and a specific PPR scheme is presented that will be studied throughout the paper. Basic properties of this PPR scheme are presented when the records are binary vectors with the Hamming distance and it is studied how to achieve a high level of privacy. However, this solution requires a significantly large number of servers. This leads us to the main problem studied in the paper in which the minimum number of servers that provide a given privacy level is investigated. Lower bounds on this value are studied in Section~\ref{sec:lower}. Upper bounds by specific constructions are given in Section~\ref{sec:construct}. The analysis of these bounds is given in Section~\ref{sec:analysis}. In Section~\ref{sec:other_metrics}, we generalize our results beyond the Hamming metric to distance regular graphs such as the Johnson graph. Finally, Section~\ref{sec:conclude} concludes the paper.

\vspace{-0.2cm}

\section{Private Proximity Retrieval Schemes and Basic Properties}
\label{sec:protocol}

\subsection{Basic Definitions}

Assume a database $\cX$ consisting of $M$ records is stored on $N$ servers, where each server stores the whole database. Assume further that a user has a record $x$ and wants to know which records in the database are similar to his record, that is, within some distance $r$ with respect to a given metric. The user wants to keep the contents of his record as private as possible, i.e., reduce the amount of information which is revealed to the servers about his record.

Let $\cV$ be a finite set and let $d$ be a metric on $\cV$, $d:\cV\times \cV\rightarrow \mathbb{R}_{\geq 0}$. For $x\in \cV$, we define by $B(x,r)$ its ball of radius $r$, that is, the set $B(x,r) \triangleq \{y\in \cV \ | \ d(x,y)\leq r\}$.  Let $X$ be a random variable on $\cV$ and let $\cX=(x_1,\ldots,x_M)$ consist of $M$ i.i.d.~samples of $X$.  We refer to $\cX$ as the \emph{database}, and assume a copy of $\cX$ is stored on each of the $N$ servers.  A user has his own independent sample $x$ of $X$, and wishes to exactly compute the set
\[
I(x,r) \triangleq \{m\in \{1,\ldots,M\} |\ x_m\in B(x,r)\}.
\]
In other words, the user wishes to answer the question ``Is $x_m\in B(x,r)$?'' for all $x_m$ in the database. We refer to $r$ as the \textbf{\emph{search radius}} of the user.

We denote by $H(\cdot)$ the binary entropy function, given by $H(p)=-p\log_2{p}-(1-p)\log_2{(1-p)}$ for $0\leq p\leq 1$. For a discrete random variable $Y$ supported on a set $\cY$, we define its entropy by
\[
H(Y) = \mathbb{E}_Y\left[\log_2\left(\frac{1}{Pr(y)}\right)\right] = -\sum_{y\in \cY}Pr(y)\log_2(Pr(y)).
\]
It will always be clear from the context whether the function $H(\cdot)$ is taking a scalar or a random variable as its argument.  For a positive integer $n$, the set $\{1,2,\ldots,n\}$ is denoted by $[n]$.

\vspace{-0.15cm}

\begin{definition}\label{def:ppr}
Let $\mathcal{X} = (x_1,\ldots,x_M)$ be a database of $M$ records sampled i.i.d.\ from a random variable $X$ on a metric space $\cV$.  Suppose that $\mathcal{X}$ is replicated on $N$ non-communicating servers, and that a user has his own record $x\in \cV$ and a desired search radius $r>0$.
A \textbf{private proximity retrieval} (\textbf{PPR}) scheme for this setup consists of three algorithms.
\begin{enumerate}
    \item A randomized algorithm $\cQ$ that forms $N$ queries, $q_1, \ldots, q_N$, depending on $X$, $M$, the user's record $x$, and the search radius $r$.
    \item An algorithm $\cA$ that calculates answers $A_n$, for $n\in[N]$, given a query $q_n$ generated by the algorithm $\cQ$, and the database $\cX$.
    \item An algorithm $\cR$ that calculates exactly the set $I(x,r)$, given $X$, $x$, $r$, the set of queries $q_1, \ldots, q_N$, and the corresponding answers $A_1, \ldots, A_N$.
\end{enumerate}
\end{definition}

\vspace{-0.15cm}

\begin{definition}
For a given PPR scheme, the user's \textbf{privacy} at a server $n\in[N]$ is defined to be \[P_n = \frac{H(X|q_n)}{H(X)}.\]
\end{definition}
The user's privacy at server $n\in [N]$ measures the fraction of bits of $x$ which is kept private from server $n$.  Note that our definitions do not demand that a PPR scheme has perfect privacy, and a major theme of this work will be the trade-off between the privacy $P_n$, the search radius $r$, and the number of servers $N$ required by the PPR schemes. Lastly we comment here that a more general definition of this problem, which takes also into account the distortion of computing the set $I(x,r)$, has been presented in a research which has
motivated the current paper (see~\cite{EGKYZ}).



\vspace{-0.3cm}

\subsection{A Simple Scheme for Hamming Space}

The main part of the current work concentrates on the case of $\cV=\F_2^L$, $d$ is the Hamming distance on $\F_2^L$, and the random variable $X$ equals to the uniform distribution on $\F_2^L$, and this is the case to which we now specify unless stated explicitly otherwise.
For two vectors $u$ and $v$ of the same length, the Hamming distance between $u$ and $v$ is denoted by $d(u,v)$ and the Hamming
weight of $u$ is denoted by $w(u)$. The support of $u$ will be denoted by $\supp(u)$.
For $s\geq 0$, let $\cW_s$ be the set of all vectors of weight exactly $s$ in $\F_2^L$, i.e.,
$\cW_s = \{ y\in \F_2^L \ | \ w (y) = s \}$.  The following proposition will motivate our approach.

\begin{prop}
\label{prop:E=0}
Given $r,s,L\geq 0$ such that $r< L$ and $x \in \F_2^L$, we have that
\begin{equation}\vspace{-0.1cm}
\label{eq:motivate}
B(x,r) = \bigcap_{z\in \cW_s}B(x+z,r+s)
\end{equation}
if and only if $r+2s+1\leq L$.

\begin{IEEEproof}
Assume first that $r + 2s + 1 \leq L$.  By translation invariance of the Hamming distance, we can assume that $x=\bzero$.
Let $y\in B(\bzero,r)$, i.e., $y$ is a vector whose Hamming weight is at most $r$.
Clearly, $y \in B(z,r+s)$ for all $z\in \cW_s$, since $d (z,y) \leq w(z) + w(y) \leq s+r$.
Hence, $B(\bzero,r) \subseteq \bigcap_{z\in \cW_s}B(z,r+s)$.

Now let $w(y)>r$, and without loss of generality we assume that the support of $y$ is exactly the first $w(y)$ coordinates. Let $z \in \cW_s$ be the vector with support on the last $s$ coordinates. We see that
\[
d(y,z)=w(y)+w(z)-2|\supp(y) \cap \supp(z)|=w(y)+s-2 \max\{0, w(y)+s-L\}.
\]
Now if $w(y)\leq L-s$, \emph{i.e.}, the vectors $y$ and $z$ share no support, it holds that $d(y,z)=w(y)+w(z)>r+s$ and hence $y \notin B(z,r+s)$. On the other hand, if $y>L-s$ it holds that $d(y,z)=w(y)+w(z)-2(w(y)+s-L)=2L-w(y)-s\geq L-s > r+s$, where the last equation follows from $L>r+2s$. Hence, $y \notin B(z,r+s)$.

For the other direction, let $B(x,r) = \bigcap_{z\in \cW_s}B(x+z,r+s)$ and assume for contradiction that
$r+2s+1 > L$. We will exhibit a vector $y$ which is in $\bigcap_{z\in \cW}B(x+z,r+s)$ but not in $B(x,r)$,
which will lead to a contradiction. Given $x$, consider the all-ones
vector ${\bf 1}$ and let $y = x + {\bf 1}$.  Clearly $y\not\in B(x,r)$ since $d(x,y) = L>r$.
One can easily verify that easily $d(x+z,y)=d(z,{\bf 1})=L-s\leq r+s$ and hence $y$ is contained
in $B(x+z,r+s)$ for all $z \in \cW_s$. Thus, $r+2s+1 \leq L$.



\end{IEEEproof}
\end{prop}

Proposition~\ref{prop:E=0} suggests that to construct a PPR scheme for retrieving $I(x,r)$, we should apply a PPR scheme wherein the user randomly picks a different translation vector $z\in \cW_s$ for each of $N = |\cW_s| = \binom{L}{s}$ servers, and sends each server the vector $x + z$.  The queries are of the form $(x+z,r+s)$ for all $z\in \cW_s$, and are transmitted to a set of servers indexed by the elements of $\cW_s$. Thus, the server corresponding to an element $z \in \cW_s$ will respond with the indices $I(x+z,r+s)$. By Proposition~\ref{prop:E=0}, we have
\begin{equation}
I(x,r) = \bigcap_{z\in \cW_s}I (x+z,r+s),
\end{equation}
and hence the user can compute $I(x,r)$ exactly. However, the number of servers $N$ for such a scheme is unacceptably large.  With the intention of reducing the number of servers but retaining the general scheme outline, we introduce the following family of PPR schemes. 

\subsection{A General PPR Scheme and Some Basic Properties}

While the PPR scheme described in the previous subsection using the set $\cW_s$ has the advantage of being easy to describe, the number of servers is $\binom{L}{s}$, which even for small values of $L$ is unreasonable.  Our general strategy for improving on this  construction will be to consider subsets $Z\subseteq \cW_s$ which satisfy the equation of Proposition \ref{prop:E=0}, but for which $|Z|\ll |\cW_s|$.  We formalize the family of PPR schemes we will consider in the following definition.

\begin{definition}\label{PPRrsZ}  Let $\mathcal{X} = (x_1,\ldots,x_M)$ be a database of $M$ records sampled i.i.d.\ from a random variable $X$ on a metric space $\cV$.  Suppose that $\mathcal{X}$ is replicated on $N$ non-communicating servers, and that a user has his own private record $x\in \cV$.  The user also has a desired search radius $r>0$, another integer parameter $s>0$, and a set of \emph{query vectors} $Z\subseteq \cW_s$ of size $N$, all of which are public knowledge.

The PPR scheme $PPR(r,s,Z)$ associated with this setup is defined to consist of the following algorithms:
\begin{enumerate}
    \item The algorithm $\cQ$ applies a uniform random permutation of $[L]$ to the coordinates of all vectors $z_n\in Z$, and then sends the query $q_n = (x + z_n,r+s)$ to server $n = 1,\ldots,N$, where $x$ is the user's record.
    \item The algorithm $\cA$ computes
    \[
    A_n = I(x+z_n,r+s) = \{m\in [M]\ |\ d(x+z_n,x_m)\leq r+s\}.
    \]
    \item The algorithm $\cR$ computes
    \[
    I(x,r) = \bigcap_{n\in [N]}A_n = \bigcap_{n\in [N]}I(x+z_n,r+s).
    \]
\end{enumerate}
\end{definition}

\begin{proposition}\label{privacy}
For the PPR scheme $PPR(r,s,Z)$, the privacy at any server $n\in[N]$ satisfies
\[
P_n  = \frac{\log_2 \binom{L}{s}}{L}.
\]
In particular, if $\sigma = s/L$ is constant with respect to $L$, then $P_n \rightarrow H(\sigma)$ as $L\rightarrow\infty$.
\end{proposition}
\begin{IEEEproof}
The coordinate permutation used in the algorithm $\cQ$ guarantees that any single server observes an element $x+z$ which is chosen uniformly at random amongst all vectors of $\F_2^L$ which are at distance $s$ from $x$, i.e., the set $x+\cW_s$.  Therefore $H(X|q_n) = \log_2|\cW_s|$.  Since $x$ is a sample of the uniform distribution on $\F_2^L$, we have $H(X) = L$ and the result holds. In case $\sigma = s/L$, computing the limit as $L\rightarrow\infty$ is straightforward and may be done using Stirling's approximation.

\end{IEEEproof}

Let $P = P_n$ for the PPR scheme $PPR(r,s,Z)$, which by Proposition~\ref{privacy} is independent of $n$.  It is not hard to show that $P < H(\sigma)$, thus approximating $P\approx H(\sigma)$ for large $L$ slightly overestimates the privacy level.  Nevertheless, to maximize privacy, one wants $\sigma$ to be as close as possible to $1/2$, so that $H(\sigma)\approx 1$.
Apart from its role in the above brief analysis of the privacy of $PPR(r,s,Z)$, the random coordinate permutation used by the algorithm $\cQ$ is largely immaterial to the analysis of the scheme, and in what follows we will largely ignore it.

From the definition of the PPR scheme $PPR(r,s,Z)$, we see that it is possible to successfully compute the desired set of indices $I(x,r)$ if the set $Z$ satisfies $I(x,r) = \bigcap_{z_n\in Z}I(x+z_n,r+s)$. Furthermore, the scheme uses $N = |Z|$ servers and obtains an asymptotic privacy level of $H(\sigma)$ where $\sigma = s/L$. Hence, for given values of $L,s,r$, the goal is to find a set $Z$ of minimum size, which translates to the number of servers required by the PPR scheme $PPR(r,s,Z)$. For the rest of the paper we study this value which will be denoted by $N(L,s,r)$. That is, $N(L,s,r)$ is the minimum size of a set $Z$ such that for all $x\in \cV$, it holds that $I(x,r) = \bigcap_{z_n\in Z}I(x+z_n,r+s)$.

\section{Bounds on the Size of PPR Codes}
\label{sec:bounds}

In this section, we analyze the quantity $N(L,s,r)$, which is
the minimum number of servers required for the PPR schemes introduced
in Definition 4. The set of query words
related to these servers form a code for PPR schemes.
We will start by giving a formal definition and an appropriate name for these codes.
Lower bounds on the size of such codes will be derived in Section~\ref{sec:lower}. Upper bounds will be provided
in Section~\ref{sec:construct} and analysis on the bounds will be discussed in Section~\ref{sec:analysis}.

Proposition~\ref{prop:E=0} has motivated the definitions
of PPR schemes and the quantity $N(L,s,r)$ in Section~\ref{sec:protocol}.
We note that the righthand side of~(\ref{eq:motivate}) reflects the elements which
are covered by intersection of balls centered at words whose distance from the record of
the user is $s$. This is
the motivation for the following definition.
For given $L$, $s$, and $r$, such that $L \geq r+2s+1$, let $C$ be a subset of $\cW_s$ such that
\begin{equation}
\label{eq:PPRIC}
B(\bzero,r) = \bigcap_{c\in C}B(c,r+s).
\end{equation}
Such a set $C$ will be called an $(L,s,r)$ \emph{Private Proximity Retrieval Intersection Covering code},
or an $(L,s,r)$ \emph{PPRIC code} in short.
Let $N(L,s,r)$ be the minimum number of codewords in such an $(L,s,r)$ PPRIC code $C$. We continue
and discuss bounds on the sizes of such codes and some properties related to their structure.
Our main interest is in bounds for which $r$ is fixed, $L \rightarrow \infty$, and $\sigma = \frac{s}{L}$ is fixed.

An important concept that will be used in our lower bounds and constructions for PPRIC codes is a word
which has a nonempty intersection with the support of each codeword of a given PPRIC code.
Given a PPRIC code $C$, a word $v \in \F_2^L$ is called
an \emph{intersection PPR word} related to $C$,
if $\supp (v) \cap \supp (c) \neq \varnothing$ for each $c \in C$.
Such a word $v$ is called a \emph{minimal intersection PPR word} (an \emph{MIPPR word} in short) if there is no other
intersection PPR word~$u$ such that $\supp (u) \subset \supp (v)$.

\begin{lemma}
\label{lem:lw_MI}
Let $r$, $s$, and $L$ be nonnegative integers, such that $L \geq 2s+r+1$, and let $C \subseteq \cW_s$ be an $(L,s,r)$ PPRIC code.
If $v$ is an MIPPR word, then $wt (v) \geq r+3$.
\end{lemma}
\begin{IEEEproof}
Assume to the contrary that $v$ is an MIPPR word with respect to $C$ and $wt (v)\leq r + 2$, and let $y\in\F_2^L$ be any word
of weight $r + 2$ such that $\supp (v) \subseteq \supp(y)$. Since $\supp (v) \cap \supp(c)\neq \varnothing$
for all $c \in C$ and $\supp (v) \subseteq \supp(y)$,
if follows that $|\supp(y)\cap\supp(c)|\geq  1$ for all $c\in C$. Hence, for any $c\in C$, we have that
$$
d(y,c) = wt(y) + wt(c) - 2|\supp(y) \cap \supp(c)| = r+2+s-2|\supp(y) \cap \supp(c)| \leq r + s
$$
and therefore, $y \in \bigcap_{c\in C}B(c,r+s)$, a contradiction to~(\ref{eq:PPRIC}) since $wt(y) > r$.
\end{IEEEproof}
The number of codewords in a PPRIC code $C$ is always an upper bound on the weight of the related
MIPPR word and hence
\begin{corollary}
\label{cor:lower_bound}
If $r$, $s$, and $L$ are nonnegative integers, such that $L \geq 2s+r+1$, then $N(L,s,r) \geq r+3$.
\end{corollary}

By Lemma~\ref{lem:lw_MI}, the weight of
an MIPPR word is at least $r+3$. This motivates constructions
and lower bounds on the sizes of $(L,s,r)$ PPRIC codes for which the weight of each MIPPR word
is exactly~$r+3$.
In general MIPPR words of $(L,s,r)$ PPRIC codes do not have to be of weight $r+3$.
For example, let $C$ be the $(L,s,r)$ PPRIC code $C$ which contains all the $\binom{L}{s}$ words of length $L$
and weight $s$. The support of any word $v$ of weight $L-s$,
have an empty intersection with the support of its binary complement $\bar{v}$,
which has weight $s$ and hence $\bar{v} \in C$.
Therefore, any MIPPR word related to $C$ has weight $L-s+1$.
Nevertheless we have the following two conjectures (a strong one and a slightly weaker one) which are intriguing and can give some
indication on related properties of PPRIC codes.

\begin{conjecture}
\label{conj:MIPPR}
Any MIPPR word of a minimal $(L,s,r)$ PPRIC code has weight $r+3$.
\end{conjecture}
\begin{conjecture}
\label{conj:minimumMIPPR}
Any MIPPR word of minimum weight, in a minimal $(L,s,r)$ PPRIC code, has weight $r+3$.
\end{conjecture}


\subsection{Lower Bounds on the Size of PPR Codes}
\label{sec:lower}

Note that by Lemma~\ref{lem:lw_MI} the weight of any MIPPR word is at least $r+3$.
If the claims in Conjectures~\ref{conj:MIPPR} and~\ref{conj:minimumMIPPR} are correct, then this lower
bound on the weight of any MIPPR word is in fact the weight of such a word.
Lemma~\ref{lem:lw_MI} also implies that we can obtain bounds on $N(L,s,r)$ by considering words
which intersect the support of every $c\in C$ (where $C$ is a PPRIC code which attains the value of $N(L,s,r)$)
in at least one coordinate. This approach was used in
the following theorem presented in~\cite{EGKYZ} which is proved by
repeatedly applying Lemma~\ref{lem:lw_MI}.

\begin{theorem}
\label{thm:lower_bound}
If $r$, $s$, and $L$ are nonnegative integers, such that $L \geq 2s+r+1$, then
\begin{equation}
\label{eq:lower_repeat}
N(L,s,r)\geq \max_{k = 0,\ldots,r+1}\left\{\left\lceil\frac{r + 3 - k}{(1-\sigma)^k}\right\rceil\right\},
\end{equation}
where $\sigma = s/L$.
\end{theorem}

One can easily verify that the values
in~(\ref{eq:lower_repeat}) are increasing with $k$, as long as $k \leq \lfloor r+4 - 1/\sigma \rfloor$.
Thus, $N(L,s,r) \geq \left\lceil\frac{r + 3 - k}{(1-\sigma)^k}\right\rceil$,
where $\sigma = s/L$ and $k=\max \{0, \lfloor r+4 - 1/\sigma \rfloor \}$.

Note that Theorem~\ref{thm:lower_bound} implies Corollary~\ref{cor:lower_bound}.
In the sequel we obtain a lower bound which also implies Theorem~\ref{thm:lower_bound}.
The new lower bound
is based on design theory and we will present first the required definitions for this lower bound.
These definitions will be also used in our constructions given in the next subsection.
We will show that our PPRIC codes are covering codes (more commonly known as covering designs or just coverings)
which have some additional constraints.

An $(n,k,t)$ \emph{covering design}, $n \geq k \geq t > 0$, is a collection of $k$-subsets (called \emph{blocks})
of an $n$-set (w.l.o.g.  $[n]$) (whose elements are called \emph{points}),
such that each $t$-subset of the $n$-set is contained in at least one block of the collection.
Let $c(n,k,t)$ denote the minimum number of blocks in such an $(n,k,t)$ covering design.
Finally, the \emph{complement} $\bar{S}$ of the design $S$ is defined by
$$
\bar{S} \triangleq \{ [n] \setminus B ~:~ B \in S \}.
$$
Similarly for a binary code $C$ of length $n$, the \emph{complement} $\bar{C}$ is the
code formed from the binary complements of all the codewords in $C$.
In other words, the codewords in $\bar{C}$ are exactly all the words whose
set of supports is
$$
\{ [n] \setminus \supp (c) ~:~ c \in C \}.
$$

Covering designs were extensively studied mainly from a combinatorial
point of view~\cite{MiMu92}. They have also found many applications, especially in
various problems related to coding theory, e.g.~\cite{EWZ95}.
Bounds on the sizes of covering designs, i.e. on $c(n,t,k)$, are very important in our discussion
as a consequence of the connection between PPRIC codes and covering designs given in the next result.

\begin{lemma}
\label{lem:eq_design}
The set of supports in the complement of an $(L,s,r)$ PPRIC code is an $(L,L-s,r+2)$ covering design.
\end{lemma}
\begin{IEEEproof}
Let $C$ be an $(L,s,r)$ PPRIC code, $S \triangleq \{ \supp(c) ~:~ c \in C \}$,
and $\bar{S}$ its complement design.
By Lemma~\ref{lem:lw_MI}, for an $(L,s,r)$ PPRIC code all MIPPR words have at least weight $r+3$.
It implies that there are no $r+2$ coordinates which contain an element from the support of each codeword.
In other words, for each subset $P \subset [L]$ of size $r+2$ there exists a codeword $c \in C$ such that
$P \cap \supp (c) = \varnothing$.  It follows that
for each subset $P \subset [L]$ of size $r+2$ there exists a block $B \in \bar{S}$
such that $P \subset B$. Moreover,
in $\bar{S}$ each block has size $L-s$. Thus,~$\bar{S}$ forms an $(L,L-s,r+2)$ covering design.
\end{IEEEproof}

An $(n,k,t)$ \emph{Tur\'{a}n design}, $n \geq k \geq t>0$, is a collection $S$ of $t$-subsets (called \emph{blocks}) of an $n$-set,
such that each $k$-subset of the $n$-set contains at least one block of $S$.
There is a simple connection between covering designs and Tur\'{a}n designs implied
by the complement of the design.
\begin{theorem}
\label{thm:Co_Tu}
$S$ is an $(n,k,t)$ covering design if and only if $\bar{S}$ is an $(n,n-t,n-k)$ Tur\'{a}n design.
\end{theorem}

\begin{corollary}
\label{cor:eq_design}
The supports of an $(L,s,r)$ PPRIC code form an $(L,L-r-2,s)$ Tur\'{a}n design.
\end{corollary}

$(n,k,t)$ Tur\'{a}n designs were considered only
for small values of $k$ and $t$, i.e. $n \geq 2k$. Hence, most of the results on
Tur\'{a}n designs will not help in our exposition on PPRIC codes.
Those which can help were translated into covering designs by Theorem~\ref{thm:Co_Tu}.
Hence, we won't consider Tur\'{a}n designs in the sequel.

\begin{remark}
Lemma~\ref{lem:eq_design} and Corollary~\ref{cor:eq_design} imply that given an $(L,s,r)$ PPRIC code $C$,
for each word $v$ of length $L$ and weight ${L-r-2}$,
there exists a codeword $c \in C$ such that $\supp(c) \subset \supp(v)$.
For each word $v$ of length $L$ and weight ${L-r-2-2\gamma}$, $\gamma \geq 0$,
there exists a codeword $c \in C$ such that $|\supp(c) \cap \supp(v)| \geq s - \gamma$.
This is exactly what is required for data compression as explained in~\cite{EWZ95},
where some bounds including asymptotic bounds in this direction are given.
\end{remark}

An $(L,s,r)$ PPRIC code $C$ is the complement of an $(L,L-s,r+2)$ covering design. Therefore,
for every word $v$ of weight $r+1$ or $r+2$, there exists a codeword $c\in C$ such that $\supp(c) \cap \supp(v) =\varnothing$.
However, PPRIC code has some more requirements. Equality (\ref{eq:PPRIC}) suggests that, for each word $v$ of weight
$r+2\gamma-1$ or $r+2\gamma$, $\gamma \geq 1$, there exists a codeword $c\in C$ such that $|\supp(c)\cap\supp(v)|<\gamma$. This implies that $d(v,c)=wt(v)+s-2|\supp(c)\cap\supp(v)| \geq r+2\gamma-1+s-2(\gamma-1)>r+s$ as expected.

These properties of PPRIC codes imply a method to verify whether a code $C$ is indeed an $(L,s,r)$ PPRIC code.
To show that $C$ is a PPRIC code, we just need to prove that any word of length $L$ which intersects each codeword of $C$ in
at least $\gamma$ coordinates should have weight at least $r+2\gamma+1$. Equivalently, to show that $C$ is not
a PPRIC code, we just need to find one word of weight $r+2\gamma$ which intersects each codeword of $C$ in at least $\gamma$ coordinates, and therefore has distance at most $r+s$ from all the codewords of $C$.

Lemma~\ref{lem:eq_design} also implies the following important lower bound.
\begin{corollary}
\label{cor:eq_design_bound}
If $r$, $s$, and $L$ are nonnegative integers, such that $L \geq 2s+r+1$, then $N(L,s,r) \geq c(L,L-s,r+2)$.
\end{corollary}

Corollary~\ref{cor:eq_design_bound} implies several lower bounds on $N(L,s,r)$.
In fact, most of our lower bounds are based on the lower bounds on the size of the related covering designs, but
there are some parameters for which $N(L,s,r) > c(L,L-s,r+2)$ as we will see in the sequel.
The first lower bound is the basic covering bound.
\begin{lemma}
\label{lem:lw_covering}
If $r$, $s$, and $L$ are nonnegative integers, such that $L \geq 2s+r+1$, then
$$
N(L,s,r) \geq \frac{\binom{L}{r+2}}{\binom{L-s}{r+2}}~.
$$
\end{lemma}

The next result is well known as the Sch\"{o}nheim's bound~\cite{Sch64}.
\begin{lemma}
\label{lem:Sch}
If $n>k>t>0$, then
$$
c(n,k,t) \geq \left\lceil \frac{n}{k} c(n-1,k-1,t-1) \right\rceil ~.
$$
\end{lemma}

Lemma~\ref{lem:Sch} can be applied iteratively to obtain several results.
\begin{corollary}
\label{cor:Sch_it_part}
If $n>k>t>0$, then
$$
c(n,k,t) \geq \left\lceil \frac{n}{k} \left\lceil \frac{n-1}{k-1} \cdots  \left\lceil \frac{n-\ell+1}{k-\ell+1} c(n-\ell,k-\ell,t-\ell) \right\rceil \cdots \right\rceil \right\rceil~.
$$
\end{corollary}
\begin{corollary}
\label{cor:Ser_it_part}
If $r$, $s$, and $L$ are nonnegative integers, such that $L \geq 2s+r+1$, then
$$
N(L,s,r) \geq \left\lceil \frac{L}{L-s} \left\lceil \frac{L-1}{L-s-1} \cdots  \left\lceil \frac{L-\ell+1}{L-s-\ell+1} c(L-\ell,L-s-\ell,r+2-\ell) \right\rceil \cdots \right\rceil \right\rceil~.
$$
\end{corollary}
\begin{corollary}
\label{cor:Sch_it}
If $n>k>t>0$, then
$$
c(n,k,t) \geq \left\lceil \frac{n}{k} \left\lceil \frac{n-1}{k-1} \cdots  \left\lceil \frac{n-t+1}{k-t+1} \right\rceil \cdots \right\rceil \right\rceil~.
$$
\end{corollary}
Corollary~\ref{cor:Sch_it} implies Lemma~\ref{lem:lw_covering}.
It also implies that $c(n,k,t) \geq t+1$
and by applying Corollary~\ref{cor:Ser_it_part}
we obtain Theorem~\ref{thm:lower_bound}.

Usually, $(n,k,t)$ covering designs were considered
for small~$k$ and~$t$, i.e. $n \geq 2k$. In our context we need small~$t$ and large $k$, i.e.
small $n-k$. These types of designs were considered for a given $m$ and $t$, and
pairs $(n,k)$ such that $m=c(n,k,t)$. There has been extensive research in this direction,
which is exactly what is required for our bounds on PPRIC codes. The main difficulty as was
pointed in~\cite{MiMu92} is to establish the correct lower bound. This usually has required a very long and complicated
proof.
Mills~\cite{Mil79} and Todorov~\cite{Tod85} found the maximum ratio $\frac{n}{k}$ on the pairs $(n,k)$ such that
$m=c(n,k,t)$. The first value was given by Mills~\cite[Theorem 2.2 and Theorem 2.3]{Mil79}.
This is quoted as in Todorov~\cite[Theorem 2]{Tod85}.

\begin{theorem}
\label{thm:Mills}
Let $n>k>t>0$ and assume that $m=c(n,k,t)$.
If $m$ and $t$ are positive integers such that $t < m \leq \frac{3(t+1)}{2}$, then
$$
\frac{n}{k} \leq \frac{3t+3-m}{3t+1-m} ~.
$$
Therefore, if $\frac{n}{k}\le \frac{t+1}{t}$ then $c(n,k,t)\ge t+1$ ; If $t+1<m\le \frac{3(t+1)}{2}$ and $\frac{3t+4-m}{3t+2-m} <\frac{n}{k}\le \frac{3t+3-m}{3t+1-m}$, then ${c(n,k,t)\ge m}$.
\end{theorem}

Clearly, we have to translate the results from the pair of parameters $(n,k)$ to the
pair $(L,s)$ which is performed by the simple equation $\frac{n}{k} = \frac{L}{L-s}$.
This implies that $\frac{k}{n} = \frac{L-s}{L} = 1 - \frac{s}{L}$ and $\frac{s}{L} = 1 - \frac{k}{n}$.

\begin{corollary}
\label{cor:Mills}
Let $r$, $s$, and $L$ be nonnegative integers, such that $L \geq 2s+r+1$.
If $m$ and $r$ are nonnegative integers such that $r+2 < m \leq \frac{3(r+3)}{2}$, then
$$
\frac{L}{s} \geq \frac{3r+9-m}{2} ~.
$$
If $\frac{L}{s} \geq r+3$ then $N(L,s,r)\ge r+3$;
If $r+3<m\le \frac{3(r+3)}{2}$ and $\frac{3r+9-m}{2} \leq \frac{L}{s} <\frac{3r+10-m}{2}$, then $N(L,s,r)\ge m$.
\end{corollary}

Theorem~\ref{thm:Mills} was extended by Todorov~\cite{Tod85} as follows.

\begin{theorem}
\label{thm:Tod1}
If $n>k>t>0$ and $m=c(n,k,t)$, then for
every integer $\ell \geq 1$ we have the following results.
\begin{enumerate}
\item If $m=3\ell +1$ and $t=2\ell -1$, then $\frac{n}{k} \leq \frac{9\ell-1}{9\ell-7}$.
\item If $m=3\ell +2$ and $t=2\ell$, then $\frac{n}{k} \leq \frac{6\ell+3}{6\ell-1}$.
\item If $m=3\ell +3$ and $t=2\ell$, then $\frac{n}{k} \leq \frac{3\ell+1}{3\ell-1}$.
\end{enumerate}
\end{theorem}

Theorem~\ref{thm:Mills} and Theorem~\ref{thm:Tod1} lead to the following consequences.
\begin{corollary}
\label{cor:Tod1}
If $n>k>t>0$ and $m=c(n,k,t)$, then for
every integer $\ell \geq 1$ we have the following results.
\begin{enumerate}
\item If $t$ is odd and $\frac{3t+3}{3t-1}<n/k\le\frac{9t+7}{9t-5}$, then $c(n,k,t)\ge \frac{3t+5}{2}$.

\item If $t$ is even and $\frac{3t+4}{3t}<n/k\le\frac{3t+3}{3t-1}$, then $c(n,k,t)\ge \frac{3t+4}{2}$.

\item If $t$ is even and $\frac{3t+3}{3t-1}<n/k\le\frac{3t+2}{3t-2}$, then $c(n,k,t)\ge \frac{3t+6}{2}$.
\end{enumerate}
\end{corollary}

\begin{corollary}
\label{cor:Todorov}
Let $r$, $s$, and $L$ be nonnegative integers, such that $L \geq 2s+r+1$.
\begin{enumerate}
\item If $r$ is odd and $\frac{9r+25}{12} \leq L/s < \frac{3r+9}{4}$, then $N(L,s,r) \geq \frac{3r+11}{2}$.
\item If $r$ is even and $\frac{3r+9}{4} \leq L/s < \frac{3r+10}{4}$, then $N(L,s,r) \geq \frac{3r+10}{2}$.
\item If $r$ is even and $\frac{3r+8}{4} \leq L/s < \frac{3r+9}{4}$, then $N(L,s,r) \geq \frac{3r+12}{2}$.
\end{enumerate}
\end{corollary}

So far we have listed several lower bounds of $N(L,s,r)$ implied by lower bounds
on $c(n,k,t)$. In the sequel it will be shown that
equality holds for almost all the bounds in Corollaries~\ref{cor:Mills} and~\ref{cor:Todorov},
by presenting an optimal covering design whose complement
is a PPRIC code. However, there is an exception
for Corollary~\ref{cor:Todorov}(3) when $r=0$, which is the next result,
demonstrating the difference between the lower bounds on $c(n,k,t)$ and $N(L,s,r)$.
The proof is quite tedious and to simplify some parts we present some auxiliary
claims throughout the proof.

\begin{theorem}
\label{thm:best_bound}
If $L$ and $s$ are integers such that $L/s=2+\epsilon$, where $\epsilon<\frac{1}{8}$, then $N(L,s,0)>6$.
\end{theorem}

\begin{IEEEproof}
Assume that for $r=0$ and $L/s=2+\epsilon$, where $\epsilon<\frac{1}{8}$,
there exists an $(L,s,r)$ PPRIC code with only $6$ codewords $A,B,C,D,E,F$.
If some five codewords have a nonempty intersection then obviously we can find two coordinates
intersecting all codewords, a contradiction to Lemma~\ref{lem:lw_MI}.
We distinguish now between four cases depending on the number of four codewords
with a nonempty intersection. The distinction is between three
such quadruples of intersection codewords (called later \emph{intersecting quadruples}
and similarly we have \emph{intersecting triples}), exactly two such quadruples,
exactly one such quadruple, or no such quadruple of codewords.
Note that if there exist two such quadruples $\{ A,B,C,D \}$ and $\{ A,B,E,F \}$
then there exists an MIPPR word with weight two contradicting Lemma~\ref{lem:lw_MI}.
Therefore, any two intersecting quadruples should have exactly three codewords in common.
W.l.o.g. two of the quadruples are $\{ A,B,C,D \}$ and $\{ A,B,C,E \}$, a third quadruple can
be w.l.o.g. either $\{ A,B,C,F \}$ or $\{A,B,D,E\}$. It is easy to verify that in both cases there is no
other quadruple which has three codewords in common with the other three quadruples. This implies
that the four cases which follow are comprehensive. We continue with a lemma which will be
used throughout the proof. A \emph{type of intersecting triples} will be characterize by the
codewords which participate in the nonempty intersection. Similarly \emph{type of intersecting quadruples}
is characterized.

\begin{lemma}
\label{lem:2t}
There do not exist $2t$ types of intersecting triples, where each codeword is contained in
exactly $t$ types of intersecting triples.
\end{lemma}
\begin{IEEEproof}
Let $v$ be a word of weight $2t$ which intersects each type in one coordinate. The weight of $v$ is $2t$ and
it intersects each codeword in $t$ coordinates. Hence, for $X \in \{ A,B,C,D,E,F\}$ we have
$d(v,X) = 2t + s -2t=s$, a contradiction to~(\ref{eq:PPRIC}).
\end{IEEEproof}
Now, we continue with the four cases.

\noindent
{\bf Case 1:} There are exactly three such intersecting quadruples.

W.l.o.g. the first two quadruples are $\{ A,B,C,D \}$ and $\{ A,B,C,E \}$,
i.e. $A \cap B \cap C \cap D \neq \varnothing$ and $A \cap B \cap C \cap E \neq \varnothing$.
W.l.o.g., a third quadruple can be either $\{ A,B,C,F \}$ or $\{ A,B,D,E \}$.

\begin{enumerate}
\item If $A \cap B \cap C \cap F \neq \varnothing$, then $D$, $E$, $F$ should be pairwise disjoint
(otherwise there exists an MIPPR word with weight two, contradicting Lemma~\ref{lem:lw_MI}).
But, this implies that $|D \cup E \cup F| \geq 3s > L$, a contradiction.

\item If $A \cap B \cap D \cap E \neq \varnothing$, then to avoid an MIPPR word with weight two $F$
should be disjoint from $C \cup D \cup E$ and hence
$[L]$ is a union of three pairwise disjoint sets, $F$, $C \cap D \cap E$ and $R$.
If $|C \cap D \cap E| =\beta s$, then it implies that
$$
3s = |C|+|D|+|E| \leq 3|C \cap D \cap E| + 2|[L] \setminus (F \cup (C \cap D \cap E))| = 3 \beta s +
2(1+\epsilon-\beta)s=(2+2\epsilon+\beta)s~,
$$
and hence $1 - 2\epsilon \leq \beta$, i.e. $|C \cap D \cap E| \geq (1 - 2\epsilon)s$. Therefore, we have
$$
|R|=| [L] \setminus (F \cup (C \cap D \cap E))| \leq (2+\epsilon)s - s- (1 - 2\epsilon)s = 3\epsilon s .
$$

Clearly to avoid more intersecting quadruples, $A$ is disjoint to $C \cap D \cap E$ and also
$B$ is disjoint to $C \cap D \cap E$. This implies that $|A \cap R| \leq 3 \epsilon s$ and $|A \cap F| \geq (1 - 3\epsilon)s$
and similarly $|B \cap F| \geq (1 - 3\epsilon) s$ and hence $|A \cap F| + |B \cap F| \geq (2 - 6 \epsilon)s > s = |F|$ and
therefore $A \cap B \cap F \neq \varnothing$. Since also $C \cap D \cap E \neq \varnothing$, it follows that there exists
an MIPPR word with weight two, contradicting Lemma~\ref{lem:lw_MI}.
\end{enumerate}

\noindent
{\bf Case 2:} There are exactly two such quadruples of codewords.

W.l.o.g. the two quadruples are $\{ A,B,C,D \}$ and $\{ A,B,C,E \}$. To avoid an MIPPR word with weight two,
$F$~has to be disjoint from $D \cup E$.
Let $|D \cap E| = \beta s$ and since $F$ is disjoint from $D \cup E$,
it follows that $D \cup E \subseteq [L] \setminus F$.  This implies that
$$
2s = |D|+|E| \leq 2|D \cap E| + |([L] \setminus (F \cup (D \cap E))| = 2 \beta s +
(1+\epsilon-\beta)s=(1+\epsilon+\beta)s~,
$$
and hence $1 - \epsilon \leq \beta$, i.e. $|D \cap E| \geq (1 - \epsilon)s >0$. Therefore, we have
$$
| [L] \setminus (F \cup (D \cap E))| \leq (2+\epsilon)s - s- (1 - \epsilon)s = 2\epsilon s .
$$

Let $[L]$ be the union of three pairwise disjoint sets, $F$, $D \cap E$ and $R$.
Assume first that $X\cap Y\cap F = \varnothing$, where $\{X,Y\}\subset\{A,B,C\}$. It implies that
$$
|A|+|B|+|C| \leq |A \cap F| + |B \cap F| + |C \cap F| + |A \cap D \cap E| + |B \cap D \cap E| + |C \cap D \cap E|
+ |A \cap R| + |B \cap R| + |C \cap R|
$$
$$
\leq |F| + |D \cap E| + 3 |R| = (2+\epsilon)s + 2 \cdot 2\epsilon s = 2s  +  5 \epsilon s < 3s,
$$
a contradiction. Therefore, w.l.o.g. we assume that $A\cap B \cap F \neq \varnothing$.
This implies that $C \cap D \cap E = \varnothing$ to avoid an MIPPR word with weight two and
hence $|C \cap F| = |C| - |C \cap D \cap E| - |C \cap R| \geq s - 0 - 2\epsilon s = (1-2\epsilon)s$.
We continue and distinguish between two subcases.

\noindent
{\bf Case 2.1:} If $A\cap C \cap F\neq \varnothing$, then the same arguments imply that $|B \cap F| \geq (1-2\epsilon)s$.
Hence, $|B \cap F| + |C \cap F| \geq (2-4\epsilon)s > s$ which is only possible if $B \cap C \cap F \neq \varnothing$.
Therefore, the same arguments also imply that $|A \cap F| \geq (1-2\epsilon)s$. But now,
$$
|A \cap F| + |B \cap F| + |C \cap F| \geq (3-6\epsilon)s > 2s,
$$
which implies that $A\cap B \cap C \cap F \neq \varnothing$ , contradicting the assumption that we have exactly two
distinct intersecting quadruples.

\noindent
{\bf Case 2.2:} If $C \cap F$ is disjoint from $A \cup B$, then recall that $|C \cap F| \geq (1-2\epsilon)s$
and it implies that $|A \cap F| \leq 2 \epsilon s$ and $|B \cap F| \leq 2 \epsilon s$. Hence,
$$
|A \cap ([L]\setminus F)| + |B \cap ([L]\setminus F)| + |D \cap ([L]\setminus F)| + |E \cap ([L]\setminus F)|
$$
$$
\geq (1 -2\epsilon)s + (1 -2\epsilon)s +s +s = 4s - 4\epsilon s > 3(1 + \epsilon)s.
$$
Since $|[L]\setminus F| = (1+\epsilon)s$ it follows that there exists one coordinate in $[L]\setminus F$ which
is contained in the four codewords $A$, $B$, $D$, $E$, i.e. $A \cap B \cap D \cap E \neq \varnothing$,
a contradiction to the fact that there are exactly
two intersecting quadruples.

\noindent
{\bf Case 3:} There is exactly one such intersecting quadruple of codewords. This
case is solved in Lemma~\ref{lem:one_quad}.

\noindent
{\bf Case 4:} There are no such intersecting quadruples of codewords.

Assume that there are exactly $\beta s$ coordinates which are contained in three codewords.
Hence, $6s = |A|+|B|+|C|+|D|+|E|+|F| \leq 3 \beta s+2(2+\epsilon-\beta)s$ since $(2+\epsilon-\beta)s$
coordinates are contained in at most two codewords. It implies that $\beta \geq 2-2\epsilon$.
We continue and distinguish between two cases:

\noindent
{\bf Case 4.1:} Any two codewords are contained in some intersecting triple.

The \emph{types} of candidates for these intersecting triples are as follows (the notation $\cap$ is omitted for simplicity).
       $$
       \begin{array}{cccccccccc}
         ABC & ABD & ABE & ABF & ACD & ACE & ACF & ADE & ADF & AEF \\
         DEF & CEF & CDF & CDE & BEF & BDF & BDE & BCF & BCE & BCD
       \end{array}
       $$

To avoid an MIPPR word with weight two there are no two triples from the same column in the code.
Consider now the type of intersecting triples which are contained in the code.
W.l.o.g we may assume that $A$ is a codeword which appears the most number of times in distinct types of
intersecting triples (other might appear the same number of times).
There are $\beta s$ coordinates which are contained in three codewords, where $\beta s \geq (2 - 2 \epsilon)s >s$,
and since $|A|=s$, it follows that not all the types contain $A$.
Hence, w.l.o.g we may assume the existence of the intersecting triple $D\cap E\cap F$ in the code.
Since $B$ and $C$ are involved in an intersecting triple and to avoid an MIPPR word with weight two
it cannot be $A\cap B\cap C$, it follows that w.l.o.g we may assume the existence of the intersecting triple $B\cap C\cap F$.
The updated table of candidates (where the red color is for a triple in the code and strike-through line is for a triple not in the code)
is as follows.
       $$
       \begin{array}{cccccccccc}
         \cancel{ABC} & ABD & ABE & ABF & ACD & ACE & ACF & \cancel{ADE} & ADF & AEF \\
         {\color{red}DEF} & CEF & CDF & CDE & BEF & BDF & BDE & {\color{red}BCF} & BCE & BCD
       \end{array}
       $$

Since the number of existing types of triples containing $A$ in the code is no less than that of the
types containing~$F$, we should have at least two types of triples in the code containing $A$ but not containing $F$.
The possible types are in the set$\{A\cap B\cap D,~A\cap B\cap E,~A\cap C\cap D,~ A\cap C\cap E \}$.
If the types of intersecting triples $A\cap B\cap D$ and $A\cap C\cap E$ are in the code
(or if $A\cap B\cap E$ and $A\cap C\cap D$ are in the code),
then these two triples, together with $D\cap E\cap F$ and $B\cap C\cap F$ form a contradiction to Lemma~\ref{lem:2t}.
Hence, w.l.o.g the types $A\cap B\cap D$
and  $A\cap B\cap E$ are in the code. The table of types in the code is updated as follows.
       $$
       \begin{array}{cccccccccc}
         \cancel{ABC} & {\color{red}ABD} & {\color{red}ABE} & ABF & \cancel{ACD} & \cancel{ACE} & ACF & \cancel{ADE} & ADF & AEF \\
         {\color{red}DEF} & \cancel{CEF} & \cancel{CDF} & CDE & BEF & BDF & BDE & {\color{red}BCF} & BCE & BCD
       \end{array}
       $$

Recall that each pair of codewords is part of a type which is contained in the code.
Consider the pair $\{ A , C\}$, one can readily verify that $A\cap C\cap F$ is the only possible type
which remained in the table to be in the code. The three types $A\cap C\cap F$, $D\cap E \cap F$ and $A\cap B \cap E$
forbid the appearance of the type $B\cap C\cap D$ in the code by Lemma~\ref{lem:2t}.
Finally, there is a type containing $C \cap D$ in the code and the only remaining choice is $C\cap D\cap E$.
The table for the types of triples is as follows.
       $$
       \begin{array}{cccccccccc}
         \cancel{ABC} & {\color{red}ABD} & {\color{red}ABE} & ABF & \cancel{ACD} & \cancel{ACE} & {\color{red}ACF} & \cancel{ADE} & ADF & AEF \\
         {\color{red}DEF} & \cancel{CEF} & \cancel{CDF} & {\color{red}CDE} & BEF & BDF & BDE & {\color{red}BCF} & BCE & \cancel{BCD}
       \end{array}
       $$

These six intersecting triples form a contradiction to Lemma~\ref{lem:2t}.

\noindent
{\bf Case 4.2:} There is a pair of codewords, say $E$ and $F$ which are not contained together
in any intersecting triple. This case is solved in Lemma~\ref{lem:no_quadB}.

Thus, all possible cases were analyzed and it was shown that in the given range there is no PPRIC code with
six codewords and hence $N(L,s,0) > 6$.
\end{IEEEproof}

Theorem~\ref{thm:best_bound} is important as it presents a set of parameters for which the
smallest PPRIC code has larger size than the smallest related covering design.
This will be further discussed in Section~\ref{sec:analysis}.

\begin{lemma}
\label{lem:one_quad}
If $L$ and $s$ are integers such that $L/s=2+\epsilon$, where $\epsilon<\frac{1}{8}$, then an $(L,s,0)$ PPRIC code
with six codewords $A,B,C,D,E,F$ and exactly one intersecting quadruple does not exist.
\end{lemma}

\begin{IEEEproof}
W.l.o.g, the intersecting quadruple is $A\cap B\cap C\cap D$, which implies that $E$ and $F$ are disjoint,
and hence $|A\cap B\cap C\cap D| = \pi s \leq \epsilon s$.
Each one of the $2s$ coordinates of $E \cup F$ can support at most three codewords. Those
which support exactly three codewords will be called \emph{saturated}.
Assume $E \cup F$ contains $\beta s$ saturated coordinates and therefore each of the other $(2-\beta)s$ coordinates
are contained in at most two codewords and hence they are unsaturated coordinates.
$|A|+|B|+|C|+|D|+|E|+|F|$ is equal $6s$ and it is also at most $4 \epsilon s + 3\beta s+2(2-\beta)s$.
Therefore, $6\leq 4\epsilon+3\beta+2(2-\beta)$ and since $\epsilon < 1/8$, it follows
that $\beta \geq 2-4\epsilon > \frac{3s}{2}$.
We continue to show, by using other counting arguments, that the number of saturated coordinates in the code is
smaller than $(2-4\epsilon)s$, which is a contradiction.

For the saturated coordinates we only have to consider intersecting triples which have the form
$\{ X,Y,Z \}$, where $X,Y \in \{A,B,C,D\}$ and $Z \in \{E,F\}$. There are twelve such possible intersecting
triples, but to avoid an MIPPR word with weight two if $\{X_1,Y_1,Z_1\}$ is an intersecting triple,
it follows that $\{X_2,Y_2,Z_2\}$, where $\{X_1,Y_1,Z_1\}\cap \{X_2,Y_2,Z_2\} = \varnothing$,
is not an intersecting triple, and hence at most six such intersecting triples can be contained in the code. W.l.o.g. assume that
$A \cap B \cap E$ is such an intersecting triple with the largest size and that ${|A\cap B \cap E|=\gamma s}$. Since there are
only six possible intersecting triples and their total size is greater than $\frac{3s}{2}$, it follows that
$6 \gamma s >  \frac{3s}{2}$ and hence $\gamma > \frac{1}{4}$.

If $C\cap D\cap E =\varnothing$, then we consider the distribution of $C$ and $D$ in the code. First note that since
${|A\cap B \cap E| > \frac{1}{4} s}$, it follows that $|E \cap C| + |E \cap D| < \frac{3s}{4}$. Moreover, to avoid an MIPPR
word with weight two $C\cap D\cap F =\varnothing$ which implies that $|F \cap C| + |F \cap D| \leq s$. Therefore,
the distribution of $C$ and $D$ on $[L]$ partitioned by $E$, $F$, and $R$ is
$$
2s=|C|+|D| \leq |E \cap C| + |E \cap D| + |F \cap C| + |F \cap D| + |R \cap C| + |R \cap D| < \frac{3s}{4} + s + 2\epsilon s < 2s,
$$
a contradiction. Hence, the code has the intersecting triples $A\cap B \cap E$ and $C\cap D\cap E$
which implies that $A\cap B \cap F=\varnothing$ and $C\cap D\cap F=\varnothing$.

Clearly, there is at least one intersecting triple which contains $F$ and w.l.o.g. we assume that
$A \cap C \cap F \neq \varnothing$. Assume further that $A \cap C \cap F $ is the
only intersecting triple which contains $F$. Since there are more than $\frac{3s}{2}$ saturated coordinates and at most $s$
of them contain $E$, it follows that $|A \cap C \cap F| > \frac{s}{2}$. But, $|A \cap B \cap E| \geq |A \cap C \cap F|$
which implies that $|A| \geq |A \cap B \cap E| + |A \cap C \cap F| > s$, a contradiction. Hence, there is at
least one more intersecting triple which contains $F$. We already have the intersecting triples $A\cap B \cap E$, $C\cap D\cap E$,
and $A \cap C \cap F$, and by Lemma~\ref{lem:2t} $B \cap D \cap F$ cannot be an intersecting triple.
Hence, w.l.o.g. a second intersecting triple which contains $F$ is $B \cap C \cap F$.

Since $|A\cap B \cap E|=\gamma s$, it follows that $|D \cap E| \leq (1-\gamma) s$ and
since $|A\cap B\cap C\cap D| \leq \epsilon s$ it follows that $|D \cap (E \cup F)| \geq (1-\epsilon)s$,
which implies that $|D \cap F| \geq (\gamma - \epsilon)s$.

By Lemma~\ref{lem:2t} the code cannot contain the intersecting triples $A\cap D \cap F$, $B\cap D \cap F$, and since
also $C\cap D \cap F$ is not an intersecting triple, it follows that all coordinates in $D \cap F$ are not saturated.
Recall, that the number of saturated coordinates is $\beta s > \frac{3s}{2}$ and hence $|D \cap F| < s/2$,
which implies that $(\gamma - \epsilon)s < s/2$ and hence $\gamma < 5/8$. Hence, there are more than $\frac{11s}{8}$
coordinates in $E \cup F$ which are not in $A \cap B \cap E$.

To summarize, we have four intersecting triples, $A \cap B \cap E$ for which we proved that $|A \cap B \cap E| < \frac{5s}{8}$,
$C \cap D \cap E$, $A \cap C \cap F $, $B \cap C \cap F$; six intersecting triples were excluded from the code,
and the two intersecting triples $A \cap C \cap E$ and $B \cap C \cap E$ may or may not be contained in the code. Note, that the last five
intersecting triples contain $C$. We call the coordinates in $E \cup F$ which support the codeword $C$, \emph{$C$-coordinates}.
All the coordinates in $E \cup F$ which are not in $A \cap B \cap E$ are $C$-coordinates or unsaturated coordinates.
Clearly, the number of $C$-coordinates is at most $(1 - \pi)s$. The number of appearances of $A$, $B$, $C$, and $D$, in the coordinates
of $E \cup F$ is at least $4s - 4\pi s - 3(\epsilon - \pi)s= (4 - 3\epsilon - \pi)s$ and hence the number of saturated
coordinates is at least $(2 - 3\epsilon - \pi)s$ (since $|E \cup F|=2s$). This implies that the number of coordinates (inside $E \cup F$)
which are unsaturated or $C$-coordinates is at most $(3\epsilon + \pi)s + (1 - \pi)s < \frac{11s}{8}$. But, these are exactly
the coordinates in $E \cup F$ which are not in $A \cap B \cap E$. There are more than $\frac{11s}{8}$ such coordinates, a contradiction.
\end{IEEEproof}

\begin{lemma}
\label{lem:no_quadB}
If $L$ and $s$ are integers such that $L/s=2+\epsilon$, where $\epsilon<\frac{1}{8}$, then an $(L,s,0)$ PPRIC code
with six codewords $A,B,C,D,E,F$, with no intersecting quadruple, where
$\{E,F\}$ is not contained in any intersecting triple, does not exist.
\end{lemma}

\begin{IEEEproof}
We define the \emph{saturated} coordinates to be the ones on which there are intersecting triples.
There are a total of $(2+\epsilon)s$ coordinates and hence at least $6s-2(2+\epsilon)s > \frac{7s}{4}$
saturated coordinates which implies that there are at most $(2+\epsilon)s - \frac{7s}{4} < \frac{3s}{8}$
unsaturated coordinates.

Since there is no intersecting triple of the form $X \cap E \cap F$, it follows that there are sixteen possible intersecting
triples as follows
       $$
       \begin{array}{cccccccc}
         ABE & ACE & ADE & BCE & BDE & CDE & ABC & ABD \\
         ABF & ACF & ADF & BCF & BDF & CDF & ACD & BCD
       \end{array}
       $$

\noindent
{\bf Claim.}
If $|X\cap Y\cap Z|=\lambda s$, then $\lambda \leq \frac{1}{2}+\epsilon$.

\noindent
To verify the claim, we note that if $\{ X, Y, Z \} \cup \{ X', Y', Z' \} = \{ A,B,C,D,E,F \}$, then
$X',Y',Z'$ are distributed only on $(2 + \epsilon - \lambda)s$ coordinates.
If $3s > 2(2 + \epsilon - \lambda)s$ then $\{ X', Y', Z' \}$ is an intersecting triple,
contradicting Lemma~\ref{lem:lw_MI}. Hence, $3s \leq 2(2 + \epsilon - \lambda)s$ implying
that $\lambda \leq \frac{1}{2}+\epsilon$.

Assume that there are no two intersecting triples of the form $X \cap Y \cap P$
and $Z \cap W \cap P$, where $P\in\{E,F\}$, $\{X,Y,Z,W\}=\{A,B,C,D\}$.
Then w.l.o.g there are only two possibilities:

\begin{enumerate}
\item the intersecting triples are chosen from $\{ ABE,ACE,ADE,ABF,ACF,ADF,ABC,ABD,ACD,BCD \}$ such that nine triples contain $A$
and only $B \cap C \cap D$ does not contain $A$. By the Claim $|B \cap C \cap D| \leq (\frac{1}{2}+\epsilon)s$,
and all the intersecting triples which contain $A$
have size at most $s$, and hence the total size of these triples is at most $(1 + \frac{1}{2}+\epsilon)s$ which is smaller than $\frac{7s}{4}$, a contradiction.

\item the intersecting triples are chosen from $\{ABE,ACE,BCE,ABF,ACF,BCF,ABC,ABD,ACD,BCD\}$.
Each triple has at least one pair from the set $\{ A,B,C \}$.
Since $|A|+|B|+|C|=3s$ and each intersecting triple have a pair from these three codewords,
it follows that the total size of these triples is at most $\frac{3s}{2}$
which is smaller than $\frac{7s}{4}$, a contradiction.
\end{enumerate}

Hence, w.l.o.g. there are at least two such intersecting triples, say $A \cap B \cap E$ and $C \cap D \cap E$.
We can also assume w.l.o.g. that $A \cap C \cap F$ is an intersecting triple.
If there is no other intersecting triples which contain $F$, then by the claim the number of saturated coordinates
which support~$F$ is at most $\frac{1}{2}+\epsilon$, and hence the total number
of unsaturated coordinates supporting $F$ is at least $\frac{1}{2}-\epsilon$ which is larger
than $\frac{3s}{8}$, a contradiction.

Hence, there is another intersecting triple which contains $F$. By Lemma~\ref{lem:2t}, $B \cap D \cap F$ is
not an intersecting triple and hence w.l.o.g. $B \cap C \cap F$ is an intersecting triple.
To summarize, we have the intersecting triples $\{ ABE, CDE, ACF, BCF \}$. Now, any triple from the
set $\{ ACE,BCE,ABC,ABD,ACD,BCD \}$ can be also an intersecting triple.
All the other triples cannot be intersecting triples.

Let $|C \cap D \cap E| = \gamma s$; note that every other intersecting triple has two codeword from
the set $\{ A,B,C \}$. Hence, the number of saturated coordinates is at most $(\gamma + \frac{3-\gamma}{2})s$.
Since the total number of saturated coordinates is at least $(2-2\epsilon)s$ it follows that
$(\gamma + \frac{3-\gamma}{2})s \geq (2-2\epsilon)s$, i.e. $\gamma > 1 -4\epsilon$.
All the coordinates in $F$ which are not contained in
$A \cap C \cap F$ or $B \cap C \cap F$ are unsaturated. The total number of
unsaturated coordinates is at most $3\epsilon s$ and hence $|A \cap C \cap F| + |B \cap C \cap F| \geq (1-3\epsilon)s$.
Therefore, we have that the total number of coordinates which support $C$ is at least
$$
|C \cap D \cap E| + |A \cap C \cap F| + |B \cap C \cap F| \geq (2 - 7\epsilon)s >s,
$$ a contradiction.

Thus, the lemma is proved.
\end{IEEEproof}

\subsection{Upper Bounds on the Size of PPRIC Codes}
\label{sec:construct}

In this subsection we present a few constructions
for PPRIC codes which imply upper bounds on $N(L,s,r)$.
Some of these bounds are already known from the constructions of covering designs in~\cite{Mil79,GPK95},
where the covering designs can be proven to be also PPRIC codes. But, these constructions of covering designs
are not always PPRIC codes, and for many parameters no constructions are given.
Other bounds known from the general constructions on covering designs are given in~\cite{Gor95}.
Analysis of our construction, and especially with comparison of related covering designs, will be given in Section~\ref{sec:analysis}.
By Proposition~\ref{prop:E=0} we know that for an $(L,s,r)$ PPRIC code, $L \geq 2s+r+1$. There
are only trivial requirements on the parameters of an $(n,k,t)$ covering design. By Corollary~\ref{cor:eq_design_bound} we have that
$N(L,s,r) \geq c(L,L-s,r+2)$ and hence for the related $(L,L-s,r+2)$ covering design we only have
$L \geq L-s \geq r+2$, i.e. $L \geq s+r+2$. The first result implies
that if $L \geq (r+3)s$, then the optimal $(L,s,r)$ PPRIC code is derived in a trivial way.

\begin{theorem}
\label{thm:r+3}
If $r$, $s$, and $L$ are nonnegative integers such that $L \geq 2s+r+1$ and
$L \geq (r+3)s$, then ${N(L,s,r) \leq r+3}$.
\end{theorem}
\begin{IEEEproof}
Let $L \geq (r+3)s$ and $C$ be a code which consists of $r+3$ codewords with
$r+3$ disjoint support sets. In view of the definition for an $(L,s,r)$ PPRIC code, we only
have to show that for any vector $v$ of weight $r+2\gamma-1$ or $r+2\gamma$, $\gamma>0$,
there exists a codeword $c \in C$ such that $d(v,c)>r+s$. Let $v$ be such a vector.
If $|\supp (c) \cap \supp (v)| \geq \gamma$ for each $c \in C$, then since the codewords of $C$
have disjoint support sets it follows that $wt(v) \geq (r+3)\gamma > r+2\gamma$, a contradiction.
Hence, there exists at least one codeword $c \in C$ such that $|\supp (c) \cap \supp (v)| < \gamma$.
For this codeword $c$ we have $d(v,c) = s  + wt (v) -2|\supp (c) \cap \supp (v)| > r+s$ and
therefore $C$ is an $(L,s,r)$ PPRIC code and the proof is completed.
\end{IEEEproof}

In view of Theorem~\ref{thm:r+3} and Corollary~\ref{cor:lower_bound} (also Theorem~\ref{thm:lower_bound}
with $k=0$) we have the following corollary.
\begin{corollary}
\label{cor:exactr+3}
If $r$, $s$, and $L$ are nonnegative integers such that $L \geq 2s+r+1$ and
$L \geq (r+3)s$, then ${N(L,s,r) = r+3}$.
\end{corollary}

As a consequence of Corollary~\ref{cor:exactr+3}, henceforth we only consider values of $L$ for which
${2s+r+1 \leq L < (r+3)s}$. It appears that many of the constructions of optimal covering designs or almost
optimal, are also constructions for optimal PPRIC codes or almost optimal, respectively. But, there are constructions
of covering designs which are certainly not PPRIC and also constructions of optimal covering designs
which are not PPRIC codes. These will be discussed in Section~\ref{sec:analysis}. But, the upper bounds
on $c(n,k,t)$ are not translated immediately into analog bounds on $N(n,n-k,t-2)$. We start with
bounds on $c(n,k,t)$ for which a related bound on $N(n,n-k,t-2)$ can be obtained. These bounds
can be found for example in~\cite{GPK95}.

\begin{lemma}
\label{lem:lemma1}
If $n > k > t > 0$ then $c(n+1,k+1,t) \leq c(n,k,t)$.
\end{lemma}

\begin{corollary}
\label{cor:cor1}
If $n > k > t > 0$ and $\alpha >0$ are integers, then $c(\alpha n,\alpha k,t) \leq c(n,k,t)$.
\end{corollary}

The analog of Lemma~\ref{lem:lemma1} is the following lemma which can be easily verified.

\begin{lemma}
\label{lem:lemma1PPR}
If $L$, $s$, and $r$, are nonnegative integer such that $L \geq 2s+r+1$, then $N(L+1,s,r) \leq N(L,s,r)$.
\end{lemma}

Corollary~\ref{cor:cor1} also has an analog bound on $N(L,s,r)$ as follows.

\begin{lemma}
\label{lem:lemma2PPR}
If $L$, $s$, $r$, and $\alpha$, are nonnegative integer such that $L \geq 2s+r+1$,
then $N(\alpha L,\alpha s,r) \leq N(L,s,r)$.
\end{lemma}

In contrary to Corollary~\ref{cor:cor1} which is an immediate consequence from Lemma~\ref{lem:lemma1},
Lemma~\ref{lem:lemma2PPR} is not a consequence of Lemma~\ref{lem:lemma1PPR} and a proof has to be provided.
We omit the proof which is very similar (and relatively simpler) to proofs of results given in the sequel.

We continue with a construction for PPRIC codes which is general and all our other constructions for PPRIC codes are variants
of this construction. To this end, we will need the following definition.
For a given $\ell \geq 0$ a \emph{design of type} $\ell$ is a collection of blocks of size $s$
from a set~$Q$ having the following property. For any subset $P \subseteq Q$, intersecting with each block in a subset whose size is at least $t$,
we have that $|P| \geq \ell + t$. In other words, for a subset $R \subseteq Q$ of size less than $\ell +t$, there
exists at least one block whose intersection with $R$ is smaller than $t$.

\noindent
{\bf Construction 1:}

Let $\ell_1,\dots,\ell_p$, $p \geq 2$, be nonnegative integers and let $r=\sum \ell_i +(p-2)t-1$.
Choose $p$ designs of types $\ell_1,\dots,\ell_p$ on~$p$ subsets of disjoint coordinates
of $[L]$ to form a code $C$.

\begin{theorem}
The code $C$ produced in Construction 1 is an $(L,s,r)$ PPRIC code, where $r=\sum_{i=1}^p \ell_i +(p-2)t-1$.
\end{theorem}
\begin{IEEEproof}
Let $S_i$, $1 \leq i \leq p$, be the related design of type $\ell_i$.
It is sufficient to show that for each word $v$ of weight $r+2t-1$ or $r+2t$, where $t \geq 1$, there exists a codeword $c \in C$,
such that $d(v,c) > r+s$. Since $r+2t = \sum_{i=1}^p (\ell_i +t) -1$, it follows that there exists one $i$
for which $|\supp(v) \cap S_i| < \ell_i +t$. Hence, at least for one of codeword (block) $c \in S_i$
we have $|\supp(v) \cap \supp(c)| < t$ and hence
$$
d(c,v) \geq s + r+2t-1 -2(t-1) > r+s
$$
which completes the proof.
\end{IEEEproof}

Next, we consider how to construct the best PPRIC code by using Construction 1.
For this purpose, an important upper bound on $c(n,k,t)$ is
the following theorem of Morley and van Rees~\cite{MoVR} generalized in~\cite{GPK95}.
Covering designs obtained by the related methods are referred
in that paper and also in~\cite{Gor95} as a special case of a dynamic programming construction.
\begin{theorem}
\label{thm:doubling}
If $n_1 > n_1-s > t_1 > 0$ and $n_2 > n_2-s > t_2 > 0$ are integers, then
$$
c(n_1 + n_2 , n_1+n_2 -s,t_1+t_2+1) \leq c(n_1,n_1 -s,t_1) + c(n_2,n_2 -s,t_2).
$$
\end{theorem}

Construction 1 implies a related result for PPRIC codes.

\begin{theorem}
\label{thm:doublingPPR}
If $L_1 > L_1-s > t_1 > 0$ and $L_2 > L_2-s > t_2 > 0$ are integers, then
$$
N(L_1 + L_2 , s,t_1+t_2-1) \leq c(L_1,L_1-s,t_1) + c(L_2,L_2-s,t_2).
$$
\end{theorem}

To apply Construction 1 we have to consider constructions for designs of type $\ell$.
For this we will define the concept of a superset.
Let $S$ be a collection of $\alpha$-subsets
(called \emph{blocks}) of an $n$-set, where $\alpha$ divides $s$ and $\bar{S}$ is an $(n,n-\alpha,t)$ covering design.
An \emph{$S$-superset} is a set of $\frac{ns}{\alpha}$ elements
partitioned into $n$ pairwise disjoint subsets called \emph{grain-sets}, where each grain-set has size $\frac{s}{\alpha}$.
Each $S$-superset will contribute exactly $|S|$ codewords to the related PPRIC code,
which will be a union of such $S$-supersets. Each codeword is defined
by the support of $\alpha$ distinct grain-sets related to the $|S|$ blocks of $S$. Clearly, each support
of such a codeword has weight $s$.

\begin{lemma}
If $\bar{S}$ is an $(n,n-\alpha,\ell)$ covering design, then the related $S$-superset (obtained from $S$),
where each grain-set has size $\frac{s}{\alpha}$, is a design of type $\ell$.
\end{lemma}

\begin{IEEEproof}
In the $S$-superset
there are $n$ grain-sets each one of size $\frac{s}{\alpha}$ with labels $\{1,2,\dots,n\}$.
Let $Q$ be the union of all these grain-sets.
For any block $B \in S$ whose size is $\alpha$, we form a codeword of size $s$ from the grain-sets
related to $B$.

Assume the contrary that the $S$-superset is not a design of type $\ell$,
i.e. that we have a subset $Q' \subseteq Q$ of size at most $\ell+t-1$, which intersects each block from
the $S$-superset in a subset whose size is at least $t$. We distinguish now between two cases:

\noindent
{\bf Case 1:} $Q'$ contains elements from at least $\ell$ distinct grain-sets.
Let $P \subseteq Q'$ be a subset of size $\ell$ with elements from distinct $\ell$ grain-sets.
$P$ intersects $\ell$ grain-sets with labels $\{i_1,\dots,i_{\ell}\} \subset \{1,2,\dots,n\}$.
By the definition of an $(n,n-\alpha,\ell)$ covering design, $\{i_1,\dots,i_{\ell}\}$ is
contained in some $(n-\alpha)$-subset $\bar{B} \in \bar{S}$.
Therefore the block $B \in S$ is disjoint from the grain-sets with labels $\{i_1,\dots,i_{\ell}\}$,
and hence $|B \cap Q'| \leq t-1 $, a contradiction.

\noindent
{\bf Case 2:} $Q'$ contains elements from exactly $\ell'$ distinct grain-sets, where $\ell' \leq \ell$,
with labels $\{i_1,\dots,i_{\ell'}\} \subset \{1,2,\dots,n\}$.
By the definition of an $(n,n-\alpha,\ell)$ covering design, $\{i_1,\dots,i_{\ell'}\}$ is also
contained in some $(n-\alpha)$-subset $\bar{B} \in \bar{S}$.
Therefore the block $B \in S$ is disjoint from the grain-sets with labels $\{i_1,\dots,i_{\ell'}\}$,
and hence $B \cap Q' = \varnothing$, a contradiction.

Therefore, we have proved that any subset of $Q$, which intersects each block from the $S$-superset
in a subset whose size is at least $t$, should have size at least $\ell+t$. Thus, an $S$-superset is a design of type $\ell$.
\end{IEEEproof}

For the following two constructions, which are special cases of Construction 1,
let $(\ell,1)$-superset denote an $S$-superset,
where $\bar{S}$ is the $(\ell+1,1,1)$ covering design which consists of $\ell+1$ blocks,
each one of size one.

\noindent
{\bf Construction 2 (A construction for odd $r$):}

Let $k \geq 1$ and $t$, $0 \leq t \leq \frac{r+1}{2}$, be an integer such that
$$
\frac{L}{s} \geq (\frac{r+3}{2}-t)\frac{k+1}{k} + t\frac{k+2}{k+1} = \frac{(r+3)(k+1)}{2k} - \frac{t}{k(k+1)}.
$$
Let $\{ A_i ~:~ 1 \leq i \leq t \}$ be a family of $t$ $(k+1,1)$-supersets and
$\{ A_i ~:~ t+1 \leq i \leq \frac{r+3}{2} \}$ be a family of $\frac{r+3}{2} -t$ $(k,1)$-supersets,
where the $\frac{r+3}{2}$ supersets are pairwise disjoint.
From each $(k+1,1)$-superset we form $k+2$ codewords of weight~$s$ and
from each $(k,1)$-superset we form $k+1$ codewords of weight $s$.

\begin{theorem}
\label{thm:odd_r}
Construction 2 produces an $(L,s,r)$ PPRIC code with
$$
(\frac{r+3}{2}-t)(k+1) + t (k+2) = \frac{(r+3)(k+1)}{2}+t
$$
codewords.
\end{theorem}
\begin{IEEEproof}
First, note that since
$$
L \geq (\frac{r+3}{2}-t)\frac{k+1}{k}s + t\frac{k+2}{k+1}s,
$$
it follows that it is possible to choose $\frac{r+3}{2}$ pairwise disjoint supersets as required.
The size of the code is readily derived by its definition.
In view of~(\ref{eq:PPRIC}) we only
have to show that for any vector $v$ of weight $r+2\gamma-1$ or $r+2\gamma$, $\gamma>0$,
there exists a codeword $c \in C$ such that $d(v,c)>r+s$.
Assume to the contrary, i.e., there exists such a vector $v$
for which $v \in \cap_{c \in C} B(c,r+s)$, i.e. $d(v,c) \leq r+s$ for each $c \in C$.
Since $wt(c)=s$ for each $c \in C$ and $d(v,c) \leq s+r$, it follows that $|\supp(v) \cap \supp(c)| \geq \gamma$
for each $c \in C$.
For any $(\ell,1)$-superset $S$, $\ell \in \{k,k+1\}$, $\supp(v)$ must have a non-empty intersection with at least one of its $\ell+1$ grain-sets.
The other $\ell$ grain-sets of $S$ form the support set of a codeword $c$ and hence intersects $\supp(v)$ in at least $\gamma$ elements.
Therefore, $|\supp(v) \cap S| \geq \gamma+1$ and since there are $\frac{r+3}{2}$ supersets, it follows that this argument
holds to all these supersets and hence we have
$$
r+2\gamma \geq wt(v) \geq \frac{r+3}{2}(\gamma+1),
$$
which is not possible when $\gamma>0$ and $r$ is an odd positive integer.
\end{IEEEproof}

\noindent
{\bf Construction 3 (A construction for even $r$):}

Let $k \geq 1$ and $t$, $0 \leq t \leq \frac{r}{2}$, be an integer such that
$$
\frac{L}{s} \geq (\frac{r+2}{2}-t)\frac{k+1}{k} + t\frac{k+2}{k+1} + 1 = \frac{(r+2)(k+1)}{2k} - \frac{t}{k(k+1)} + 1.
$$
Let $\{ A_i ~:~ 1 \leq i \leq t \}$ be a family of $t$ $(k+1,1)$-supersets,
$\{ A_i ~:~ t+1 \leq i \leq \frac{r+2}{2} \}$ be a family of $\frac{r+2}{2} -t$ $(k,1)$-supersets,
and $A_{(r+4)/2}$ be an $(s,0)$-superset,
where the $\frac{r+4}{2}$ supersets are pairwise disjoint.
From each $(k+1,1)$-superset we form $k+2$ codewords of weight $s$ and
from each $(k,1)$-superset we form $k+1$ codewords of weight~$s$.
To these codewords we add the last codeword of weight~$s$ whose support is disjoint to the supports of all
the other codewords.

Similarly to the proof of Theorem~\ref{thm:odd_r} one can prove the following theorem.

\begin{theorem}
\label{thm:even_r}
Construction 3 produces an $(L,s,r)$ PPRIC code with
$$
(\frac{r+2}{2}-t)(k+1) + t (k+2) + 1 = \frac{(r+2)(k+1)}{2}+t+1
$$
codewords.
\end{theorem}

The disadvantage of Construction 3 is that $s$ elements of $[L]$ are occupied by
just a single codeword and it requires just
one element from a related MIPPR word with $r+3$ elements. The other supersets occupy either $s+\frac{s}{k}$ or
$s+\frac{s}{k+1}$ elements of $[L]$ for two elements from the related MIPPR word.
Consider this single codeword and an arbitrary superset. Together they contribute three elements to an MIPPR word.
Replacing them by another type of superset, which still contributes
three elements to an MIPPR word, but occupies less elements of $[L]$,
may improve our upper bounds.
Moreover, using other types of supersets and their combinations
will yield better upper bounds and upper bounds for more parameters.

Finally, for many parameters we might find a smaller upper bound on $N(L,s,r)$ than the
one obtained via our constructions. One example is when $r=0$ and $L/s \geq \frac{17}{8}$.
We have constructed the following $(L,s,0)$ PPRIC code $C = \{ c_1, c_2 , c_3 , c_4 , c_5 , c_6 \}$
of size six implying that $N(L,s,0) =6$. We define the following six sets of intersecting pairs,
triples, and quadruples.

$A_1 \triangleq c_1 \cap c_2 \cap c_3 \cap c_4$,
$A_2 \triangleq c_1 \cap c_2 \cap c_5$,
$A_3 \triangleq c_3 \cap c_4 \cap c_5$,
$A_4 \triangleq c_1 \cap c_3 \cap c_6$,
$A_5 \triangleq c_2 \cap c_3 \cap c_6$, and
$A_6 \triangleq c_4 \cap c_6$.
Furthermore, let
$|A_1| = s/8$,
$|A_2| = 5s/8$,
$|A_3| = 3s/8$,
$|A_4| = s/4$,
$|A_5| = s/4$, and
$|A_6| = s/2$.

\begin{theorem}
\label{thm:eps8}
The code $C$ is an $(L,s,0)$ PPRIC code of size six.
\end{theorem}
\begin{IEEEproof}
Let $v$ be a word which intersects each codeword in $C$ in at least $t$ coordinates, $t \geq 0$.
For each $i$, $1 \leq i \leq 6$, let $a_i$ be the size of the intersection of $A_i$ and $v$.
Since $v$ intersects each codeword of $C$ in at least $t$ coordinates, it follows that
the following six inequalities must be satisfied.
\begin{equation}
\label{eq:ineq1}
\text{for}~c_1:~a_1 + a_2 + a_4 \geq t ,
\end{equation}
\begin{equation}
\label{eq:ineq2}
\text{for}~c_2:~a_1 + a_2 + a_5 \geq t ,
\end{equation}
\begin{equation}
\label{eq:ineq3}
\text{for}~c_3:~a_1 + a_3 + a_4 + a_5 \geq t,
\end{equation}
\begin{equation}
\label{eq:ineq4}
\text{for}~c_4:~a_1 + a_3 + a_6 \geq t,
\end{equation}
\begin{equation}
\label{eq:ineq5}
\text{for}~c_5:~a_2 + a_3 \geq t,
\end{equation}
\begin{equation}
\label{eq:ineq6}
\text{for}~c_6:~a_4 + a_5 + a_6 \geq t.
\end{equation}
Assume now that $v$ has weight at most $2t$, i.e.
\begin{equation}
\label{eq:ineq0}
a_1 + a_2 + a_3 + a_4 + a_5 + a_6 \leq 2t .
\end{equation}
Equation~(\ref{eq:ineq0}), equation~(\ref{eq:ineq5}), and equation~(\ref{eq:ineq6}) imply that $a_1=0$.
Equation~(\ref{eq:ineq0}), $a_1=0$, equation~(\ref{eq:ineq2}), and equation~(\ref{eq:ineq4}) imply that $a_4=0$.
Equation~(\ref{eq:ineq0}), $a_4=0$, equation~(\ref{eq:ineq1}), and equation~(\ref{eq:ineq6}) imply that $a_3=0$.
Equation~(\ref{eq:ineq0}), $a_1=0$, $a_4=0$, equation~(\ref{eq:ineq1}), and equation~(\ref{eq:ineq3}) imply that $a_6=0$,
contradicting~(\ref{eq:ineq4}). Therefore, $wt(v)\geq 2t+1$ which implies that $C$ is an $(L,s,0)$ PPRIC code.
\end{IEEEproof}

Theorem~\ref{thm:best_bound} and Theorem~\ref{thm:eps8} imply the following result
\begin{corollary}
\label{cor:best_bound}
If $\frac{17}{8} \leq \frac{L}{s} < \frac{9}{4}$ then $N(L,s,0)=6$ and if $\frac{L}{s} < \frac{17}{8}$ then $N(L,s,0) > 6$.
\end{corollary}

The $(L,s,0)$ PPRIC code $C = \{ c_1, c_2 , c_3 , c_4 , c_5 , c_6 \}$ should motivate
further research to find more constructions for $(L,s,r)$ PPRIC codes with $r>0$.
The related covering design presented in~\cite{Mil79} has size six, but it is not a PPRIC code.
This will be further discussed in Section~\ref{sec:analysis}.

\subsection{Analysis of the Lower and Upper Bounds}
\label{sec:analysis}

In this subsection we present a short analysis for the lower and upper bounds on $N(L,s,r)$ which were
obtained in the previous subsections. We are interested to know the gap between the
upper and the lower bound on $N(L,s,r)$ and in particular when these bounds coincide,
i.e. the exact value of $N(L,s,r)$ is known. We would like to know the range of parameters
for which the bounds in Corollary~\ref{cor:eq_design_bound} is attained, i.e. when the $(L,s,r)$ PPRIC
code of the minimum size has the same size as the minimum size of an $(L,L-s,r+2)$ covering design.
A followup question is for which range the size of the minimum PPRIC code is strictly larger?
Is there a region in which any $(L,L-s,r+2)$ covering design is also an $(L,s,r)$ PPRIC code?

By Corollary~\ref{cor:exactr+3} for $L \geq (r+3)s$ we have that $N(L,s,r)=r+3$.
The results of Mills and Todorov~\cite{Mil79,Tod85} on covering designs provide several lower bounds for $N(L,s,r)$
for certain regions of $\frac{L}{s}$, while our constructions meet some of these bounds. To be more specific:

$\bullet$ The lower bound of Corollary~\ref{cor:Mills} for odd $r$ is met by applying Construction 2 and Theorem~\ref{thm:odd_r} with $k=1$ and arbitrary $t$;

$\bullet$ The lower bound of Corollary~\ref{cor:Mills} for even $r$ is met by applying Construction 3 and Theorem~\ref{thm:even_r} with $k=1$ and arbitrary $t$;

$\bullet$ The lower bound of Corollary~\ref{cor:Todorov}(1) is met by applying Construction 2 and Theorem~\ref{thm:odd_r} with $k=2$ and $t=1$;

$\bullet$ The lower bound of Corollary~\ref{cor:Todorov}(2) is met by applying Construction 1 with $r/2$ $(2,1)$-supersets, $r >0$, and
the design of type 2 with 5 blocks coming from the
$(9,5,2)$ covering design with the blocks
$$
\{1,2,3,4,5\}, \{1,2,3,4,6\}, \{1,2,7,8,9\}, \{3,4,7,8,9\}, \{5,6,7,8,9\}.
$$

$\bullet$ The lower bound of Corollary~\ref{cor:Todorov}(3) for $r>0$ is met by applying Construction 1 with $r/2$ $(2,1)$-supersets
and the design of type 2 with 6 blocks coming from the $(4,2,2)$ covering design
with the blocks
$$
\{1,2\}, \{1,3\}, \{1,4\}, \{2,3\}, \{2,4\}, \{3,4\}.
$$

\vspace{0.2cm}

To summarize, we have the following exact bounds on $N(L,s,r)$.

\begin{theorem}
Let $r$, $s$, $L$, be nonnegative integers such that $L \geq 2s+r+1$.

1) If $\frac{L}{s} \geq r+3$ then $N(L,s,r)=r+3$.

2) If $r+3<m\le \frac{3(r+3)}{2}$ and $\frac{3r+9-m}{2} \leq \frac{L}{s} <\frac{3r+10-m}{2}$, then $N(L,s,r)= m$.

3) If $r$ is odd and $\frac{9r+25}{12}\leq \frac{L}{s} < \frac{3r+9}{4}$, then $N(L,s,r)=\frac{3r+11}{2}$.

4) If $r$ is even and $\frac{3r+9}{4} \leq \frac{L}{s} < \frac{3r+10}{4}$, then $N(L,s,r)=\frac{3r+10}{2}$.

5) If $r >0$ is even and $\frac{3r+8}{4} \leq \frac{L}{s} < \frac{3r+9}{4}$, then $N(L,s,r)=\frac{3r+12}{2}$.

6) If $r=0$ and $\frac{17}{8} \leq \frac{L}{s} < \frac{9}{4}$, then $N(L,s,0)=6$.
\end{theorem}

Given any other set of parameters $L$, $s$, and $r$, what is the best strategy to form
the smallest possible $(L,s,r)$ PPRIC code? Clearly, this is a difficult question, for which also
the answer for the related covering designs is not known. Construction 1 is our general construction
and it can be applied recursively using Theorem~\ref{thm:doublingPPR}. But, this application is not
unique, so dynamic programming (as mentioned in~\cite{GPK95}) should be used to obtain the best bound.
Note, that since our main interest is for an $(L,s,r)$ PPRIC for fixed $r$ and $\frac{L}{s}$ (note that $\sigma = \frac{s}{L}$),
where $L \rightarrow \infty$, it follows that our construction will be based on supersets.

\begin{example}
\label{exm:one7}
Suppose that $r=11$ and $\frac{L}{s}=7$. We choose $L=56$ and $s=8$ and write
$56=L=L_1+L_2 = 28+28$, i.e. $L_1=L_2=28$. Since $c(28,20,6) \leq 25$~\cite{Gor95}
it follows that $N(L,s,11) \leq 50$ for this case. Note that other covering designs can be used in this case.
In general using $(n,k,t)$ covering designs with larger $n$ would be better.
For this case $c(21,15,6) \leq 27$ yields an $(L,s,11)$ PPRIC code of size 54,
and $c(35,25,6) \leq 27$ yields an $(L,s,11)$ PPRIC code of size 54 implying
that for small covering designs we sometimes have to be careful in our choice of code
on which we apply Construction 1.
\end{example}

Is the value of $N(L,s,r)$ always equal to the one of $c(L,L-s,r+2)$? The answer is definitely no.
For example, by Theorem~\ref{thm:best_bound} and Corollary~\ref{cor:best_bound},
for $r=0$, $L/s=2+\epsilon$ where $\epsilon<\frac{1}{8}$, we have $N(L,s,r=0)>6$,
while $c(L,L-s,r+2=2)=6$~\cite{Mil79}. Moreover, for any given nonnegative integers $L$, $s$, $r$,
such that $L = 2s+r+1$, there exists an $(L,L-s,r+2)$ covering design $S$, where $\bar{S}$
is not an $(L,s,r)$ PPRIC code. Let $\cA$ and $\cB$ be two disjoint subsets of size~$2s$ and $r+1$, respectively.
$\cA \cup \cB$ will be the set of points for the design. Let $R$ be any $(2s,s,r+2)$
covering design on the points of $\cA$. The set of blocks of the design
is defined by
$$
S \triangleq \{ T \cup \cB ~:~ T \in R   \}.
$$
Clearly, the size of a block in $S$ is $s+r+1=L-s$, the number of blocks in $S$ is the same as the number of blocks in $R$,
and $S$ is a $(2s+r+1,s+r+1,r+2)$ covering design.
Clearly, the codewords in $\bar{S}$ have weight $s$. Let~$v$ be the vector of length $2s+r+1$ for which $\supp(v)=\cA$.
Clearly, the weight of $v$ is $2s$ and since $s>r$, it follows that $2s \geq r+1$.
But, $d(v,x)=s \leq r+s$ for each $x \in \bar{S}$, contradicting~(\ref{eq:PPRIC})
and hence $\bar{S}$ is not an $(L,s,r)$ PPRIC code.

A related interesting question is the value of $N(2s+r+1,s,r)$ for each $r$ and $s > r$.
It is easy to verify that for $r$ odd
$$
N(2s+r+1,s,r) \leq 2 \binom{s + \frac{r+1}{2}}{s}.
$$
For $r$ even
$$
N(2s+r+1,s,r) \leq \binom{s + \frac{r}{2}}{s} + \binom{s + \frac{r+2}{2}}{s}.
$$
The construction for these bounds is by a partition of $[2s+r+1]$ into two subsets $S_1$ and $S_2$ of size which
differ by at most one, and taking from $S_1$ and $S_2$ all the $s$-subsets.

As indicated, some of the exact values for $c(L,k,t)$ given in~\cite{Mil79,Tod85} are
also exact values for $N(L,L-k,t-2)$. The construction of the related covering designs
yields also PPRIC codes. The constructions given in~\cite{Mil79,Tod85} are special cases
(or variants) of Construction 2 and Construction 3. For any given $r$ and $\sigma =\frac{s}{L}$, there exists an integer $m_0 (r,\sigma)$
such that for all $m \leq m_0$, $N(L,s,r)=c(L,L-s,r+2)=m$, for large enough $L$.
Finding $m_0 (r,\sigma)$ as well as proving more relations between PPRIC codes and covering designs
are interesting problems for future research. The gap between the size of the optimal
PPRIC code and the related covering design is a very intriguing question for future research as well.

\section{Generalizations to Other Metrics}
\label{sec:other_metrics}

In our exposition so far we have considered the database to consist of binary words
of length $L$ and the distance measure which was taken between the words was the Hamming metric.
It appears that our framework of the PPR scheme and the PPRIC codes is not restricted for binary words
and to the Hamming scheme.
In this section we consider the PPR scheme in other metrics. Some of the basic foundations
of our setup are true to other metrics as well.
The result of Proposition~\ref{prop:E=0} (as well as many other results in this work) can be generalized
for other metrics on other spaces. In other words, the database $\cX$ need not to be a set of
binary words of length~$L$ with the Hamming distance taken as the distance measure.
The generalization of the PPRIC codes to other metrics yields interesting covering codes
in other metrics which were not examined before.
In this section we consider a general framework for some of these spaces with their related distance measures.

Let $\cV$ be a space with a distance measure $d$. By abuse of definition we call
such a pair, a $(\cV,d)$ scheme. The largest distance $L$ in the scheme is called
the \emph{diameter} of the scheme. A scheme will be called \emph{symmetric} if
\begin{itemize}
\item for any two elements $u_1,u_2 \in \cV$ such that $d(u_1,u_2)=k$,
there exists a vertex $v \in V$ such that $d(u_1,v)=L$ and $d(u_2,v)=L-k$;
\item the size of a ball with radius $r$ around a word $x \in \cV$ does not depend on $x$.
\end{itemize}
We will be interested only in symmetric schemes,
some of which are defined in  the next few paragraphs.

The \emph{Hamming scheme} $H_q(L)$ consists of all words of length $L$ over an alphabet of size $q$.
The \emph{Hamming distance} $d_H (x,y)$ between two words $x$ and $y$ of length $L$
over an alphabet with $q$ letters is the number of coordinates in which they differ, i.e.,
$d_H(x,y) \triangleq |\{ i : x_i \neq y_i \}|$.

The \emph{Johnson scheme} $J(n,L)$ consists of all $L$-subsets of an $n$-set (equivalent to binary words on length $n$ and weight $L$).
By abuse of notation we won't distinguish between the representation by words or the one by $L$-subsets.
The one which will be used will be understood from the context.
The \emph{Johnson distance} $d_J(x,y)$ between two $L$-subsets $x$ and $y$ is
half of the related Hamming distance, i.e., $d_J (x,y) \triangleq d_H(x,y)/2$ which also implies that that
$d_J(x,y) = |\supp(x) \setminus \supp(y)|$.
The Johnson schemes $J(n,L)$ and $J(n,n-L)$ are isomorphic
and hence w.l.o.g. we always assume that $2L \leq n$.

The \emph{Grassmann scheme} $\cG_q(n,L)$ consists of all $L$-dimensional subspaces
of an $n$-space over $\F_q$.
The \emph{Grassmann distance} $d_G(x,y)$ between two $L$-subspaces $x$ and $y$ is
defined by $d_G(x,y) \triangleq L - \dim (x \cap y)$.
The Grassmann schemes $\cG_q(n,L)$ and $\cG_q(n,n-L)$ are isomorphic
and hence w.l.o.g. we always assume that $2L \leq n$.
Codes with this metric have found application mainly in
error-correction for random network coding~\cite{EtVa11,KoKs08}.


The Hamming scheme, the Johnson scheme, and the Grassmann scheme (as well as some other schemes) are symmetric
(the proof can be found in\cite{MWS} as all these scheme are association schemes).
There are other interesting and important schemes in coding theory, but
we restrict ourselves to the previous metrics and ignore the others.
For more information on other metrics and schemes the interested reader is referred to~\cite{MWS}.

Our PPR schemes can be adapted to any symmetric scheme.
We will leave this claim as an exercise to the reader and only refer in
the sequel to the random permutation used in the scheme.
For example, in the Hamming scheme over an arbitrary alphabet this permutation is exactly
as in the binary Hamming scheme.

In our exposition there are three parameters ---
the radius $r$ for the record of the user, the radius $s$ from the proximity queries,
and the diameter $L$ of the scheme.
For an element $x \in \cV$, let $\cW^x_s$ denote the set of words in $\cV$ which are
at distance $s$ from $x$, i.e. $\cW^x_s \triangleq \{ z : d(z,x)=s \}$.

Given $x\in \cV$, we want to express the ball $B(x,r)$ as an intersection of some other balls.
The first step towards this goal is proved in the next result which generalizes one direction of Proposition~\ref{prop:E=0}.
\begin{proposition}
\label{prop:gen_E=0}
If $r,s,L$ are nonnegative integers such that $L \geq r+2s+1$ and $x$ is an element in a symmetric scheme $\cV$ whose
diameter is $L$, then
\begin{equation}
\label{eq:interB}
B(x,r) = \bigcap_{z\in \cW^x_s}B(z,r+s)~.
\end{equation}
\end{proposition}
\begin{IEEEproof}
Assume that $L \geq r+2s+1$ and let $y\in B(x,r)$, i.e., $d(x,y) \leq r$.
Clearly for all $z\in \cW^x_s$, by the triangle inequality $d(z,y) \leq d(z,x) + d(x,y) \leq s+r$,
which implies that $y\in B(z,r+s)$.
Hence,
$$
B(x,r) \subseteq \bigcap_{z\in \cW^x_s} B(z,r+s)~.
$$

To prove that the inclusion is actually an equality, let $y\in \cV$ be such that $d(x,y) >r$.
We will show that $y$ is not contained in $\bigcap_{z\in \cW^x_s} B(z,r+s)$. This will imply that
if $y$ is contained in $\bigcap_{z\in \cW^x_s} B(z,r+s)$ then $d(x,y) \leq r$, i.e.
$$\bigcap_{z\in \cW^x_s} B(z,r+s) \subseteq B(x,r)~.$$

Assume first, that $r < d(x,y) \leq r+s$.
Since the scheme is symmetric, $d(x,y) = r+i$, $0 < i \leq s$,
and the diameter of the scheme is $L\geq r+2s+1$, it follows
that there exists an element $z \in \cW^x_s$ such that $d(y,z)=r+i+s>r+s$.
Thus, $y \not\in B(z,r+s)$.

Assume now, that $r+s < d(x,y) \leq r+2s$.
Since the scheme is symmetric, $d(x,y) = r+s+i$, $0 < i \leq s$,
and the diameter of the scheme is $L\geq r+2s+1$, it follows
that there exists an element $z' \in \cW^x_{s-i+1}$ such that
$d(y,z')=r+2s+1$. For any $z \in \cW^x_s$ such that $d(z',z)=i-1$ we
have $d(z,y) \geq d(y,z') -(i-1) =r+2s+1 -(i-1)=r+2s-i+2 >r+s$.
Thus, $y \not\in B(z,r+s)$.

Finally, assume that $r+2s < d(x,y)$, i.e., $d(x,y) = r+2s+i$, $0 < i \leq s$. Hence, for all $z \in \cW^x_s$ we have that
$d(z,y) \geq r+2s+i-s >r+s$ and therefore, $y\not\in B(z,r+s)$.

This implies that $y \not\in \bigcap_{z\in \cW^x_s} B(z,r+s)$ and hence
$$
B(x,r) = \bigcap_{z\in \cW^x_s} B(z,r+s)~.
$$
%
\end{IEEEproof}

Note, that the other side given in Proposition~\ref{prop:E=0} for the binary Hamming scheme is not
true for all the other related schemes. It is not difficult to verify this claim and it is also left for the interested reader.





As for the binary Hamming scheme, a PPRIC code $C$ is a subset $C \subseteq \cW^x_s$ such that
$B(x,r) = \bigcap_{c\in C } B(c,r+s)$.
We continue to examine PPRIC codes in symmetric schemes.
For $L, r, s$, let us denote by $N_d(L,s,r)$ the minimum size of a PPRIC code with these parameters
in a scheme with metric $d$.
Some of our analysis for the binary Hamming scheme can generalized for the Hamming
scheme over an alphabet with $q$ letters.

\begin{theorem}
\label{thm:qHamm}
If $r,s,L$ are nonnegative integers such that $L \geq r+2s+1$, then $N_H(L,s,r) \leq N(L,s,r)$.
\end{theorem}
\begin{IEEEproof}
Let $C$ be a binary $(L,s,r)$ PPRIC code whose size is $N (L,s,r)$.
We claim that $C$ is also a PPRIC code over an alphabet of size $q>2$.
To complete the proof it is sufficient to show that for each word $y$ of length $L$, over
an alphabet with size $q$, whose weight is $r +t$, $t \geq 1$, there exists a codeword $c \in C$
such that $d_H (y,c) > r+s$. Let $y$ be such a word and $y'$ be the binary word obtained from $y$
by replacing all the nonzero elements in $y$ with \emph{ones}. Since $C$ is a binary PPRIC code,
it follows that there exists a codeword $c \in C$ such that $d(y',c) >r+s$. Hence we have
$$
d_H(y,c) \geq d(y',c) >r+s ~,
$$
and the proof of the claim is completed.
\end{IEEEproof}

Our next step is to generalize the concept of an MIPPR word for a PPRIC code over an arbitrary alphabet in the Hamming scheme.
Given a PPRIC code $C$, over an alphabet of size $q$, a word $v$ of length $L$ is called
an \emph{intersection PPR word} related to~$C$,
if for each $c \in C$ there exists a coordinate $i$ such that $v_i =c_i \neq 0$.
Such a word $v$ is called an MIPPR word if there is no intersection PPR word $u$
such that $\supp(u) \subset \supp(v)$.

\begin{lemma}
\label{lem:l_bound_servers_q}
Let $r$, $s$, and $L$ be integers, such that $L \geq 2s+r+1$, and let $C \subseteq \cW_s$ be a PPRIC code
over an alphabet with $q$ letters.
If $v$ is an MIPPR word, then $wt (v) \geq r+3$.
\end{lemma}
\begin{IEEEproof}
Assume to the contrary that $v$ is an MIPPR word with respect to $C$ and $wt (v)\leq r + 2$, and let $y$ be any vector of weight $r + 2$
such that $v_i = y_i$ if $v_i \neq 0$. Since $v$ in an MIPPR word, the definition of $y$ implies that for all $c \in C$,
$y$ has at least one nonzero entry $i$ such that $y_i=c_i$. Hence, for any $c\in C$, we have that
$$
d(y,c) \leq wt(y) + wt(c) - 2 = r+2+s-2 \leq r + s
$$
and therefore, $y \in \bigcap_{c\in C}B(c,r+s)$, a contradiction to the assumption that $C$ is a PPRIC code since $wt(y) > r$.
\end{IEEEproof}

\begin{corollary}
\label{cor:lower_r3}
For any admissible $L$, $s$, $r$, and an alphabet with at least two letters, $N_H(L,s,r) \geq r+3$.
\end{corollary}

Combining Theorem~\ref{thm:qHamm}, Corollary~\ref{cor:exactr+3}, and Corollary~\ref{cor:lower_r3} implies
\begin{corollary}
\label{cor:extreme}
If $L\geq s(r+3)$ in the Hamming scheme $H_q(L)$, then $N_H(L,s,r) = r + 3$.
\end{corollary}

Corollary~\ref{cor:extreme} implies that in the extreme case when $L \geq s(r+3)$ the size of
optimal PPRIC codes is the same for all the Hamming schemes.
This is not the case for all admissible values
of $L$, $s$, and $r$. One can verify that
\begin{theorem}
For all admissible values of $L$, $s$, and $r$, if $q$ is large enough then
$N_H(L,s,r)=r+3$ over an alphabet with $q$ letters.
\end{theorem}


Next, we discuss the Johnson scheme. We note again that the PPR scheme is defined exactly as in the Hamming scheme,
with one exception. The PPRIC code, formed by the queries, is related to a binary word of length $n$ and weight $L$.
Clearly, all such words are isomorphic and hence w.l.o.g. we can take $x$ to be the word with \emph{ones} in
the first $L$ coordinates, and it is akin to the all-zero word in the Hamming scheme. The set $\cW_s^x$ which replaces~$\cW_s$,
consists of all words of length $n$ with $L-s$ \emph{ones} in the first $L$ coordinates and $s$ \emph{ones}
in the remaining $n-L$ coordinates.
Finally, the random permutation (used for privacy) on the $L$ coordinates in the Hamming scheme is replaced by two random permutations,
one on the first $L$ coordinates and a second one on the last $n-L$ coordinates.
We continue to consider bounds on $N_J(L,s,r)$ and to establish a result which is analog to Corollary~\ref{cor:exactr+3}
and to Corollary~\ref{cor:extreme}.
The first lemma will be used in the proofs of Theorem~\ref{lem:min_ServersJ1}
and Theorem~\ref{lem:min_ServersJ2}.

\begin{lemma}
\label{lem:simple_dist}
If $x,y,z$ are three words in $J(n,L)$ such that $z = x \setminus \sigma \cup \beta$, where
$\sigma \subset x$ and $\beta \cap x =\varnothing$, then
$$
d_J (z,y) = |x \setminus y| + |\beta \setminus y| - |\sigma \setminus y|~.
$$
\end{lemma}
\begin{IEEEproof}
The claim is derived from the following sequence of equalities using the fact that $x$ and $\beta$ are disjoint
and $\sigma$ is contained in $x$:
$$
d_J (z,y) = | z \setminus y|=|(x \setminus \sigma \cup \beta) \setminus y| = |(x \cup \beta) \setminus (y \cup \sigma)|
=|x \setminus y| + |\beta \setminus y| - |\sigma \setminus y|~.
$$
\end{IEEEproof}

\begin{theorem}
\label{lem:min_ServersJ1}
If $L\geq s(2r+3)$ in the Johnson scheme $J(n,L)$, then $N_J(L,s,r) \leq 2r + 3$.
\end{theorem}
\begin{IEEEproof}
Let $x \in \cV$ and let $C=\{v_1,v_2,\ldots,v_{2r+3}\} \subseteq \cW^x_s$ be a code with the following properties:
\begin{enumerate}
\item[(p.1)]
$d_J(x,v_i) =s$ for each $1 \leq i \leq 2r+3$.

\item[(p.2)]
$d_J(v_i,v_j)=2s$ for each $1 \leq i < j \leq 2r+3$
\end{enumerate}

Such a code $C$ is easily constructed. Each $v_i$ differs from $x$ in $s$
of the first $L$ coordinates and $s$ of the last $n-L$ coordinates.
Any two distinct $v_i$'s differ from $x$ in $s$ different coordinates.
This implies that $L \geq s(2r+3)$ as given and $n-L \geq s(2r+3)$ which
is also true since $n-L \geq L$.
Next, we claim that
\begin{equation}
\label{eq:basic-eq}
B(x,r) = \bigcap_{i=1}^{2r+3}B(v_i,r+s).
\end{equation}
It is readily verified that $B(x,r) \subseteq \bigcap_{c\in C}B(z,r+s)$.
Assume now that $y \in \cV$ and $d_J(x,y) \geq r+1$. To
complete the proof it is sufficient to prove that $y \notin \bigcap_{c\in C}B(c,r+s)$.

Property (p.1) implies that $|x \setminus v_i|=s$ and $|v_i \cap ([n] \setminus x) |=s$. Property
(p.2) implies that for $1 \leq i < j \leq 2r+3$ we have that $|(x \setminus v_i) \cup (x \setminus v_j)|=2s$
and $|(v_i \cap ([n] \setminus x)) \cup (v_j \cap ([n] \setminus x))|=2s$.

Let $a_i \deff |y \cap (x \setminus v_i)|$ be the number of elements shared by $x$ and $y$,
which are not contained in $v_i$. Let $b_i \deff |y \cap v_i \cap ([n] \setminus x)|$ be the number of elements
shared by $y$ and $v_i$ which are outside $x$. Now, we have

$$
\sum_{i=1}^{2r+3} b_i = \sum_{i=1}^{2r+3} |y \cap v_i \cap ([n] \setminus x)|
= |y \cap \bigcup_{i=1}^{2r+3} (v_i \cap ([n] \setminus x)) |   \leq d_J(x,y) ,
$$
and similarly
$$
\sum_{i=1}^{2r+3} a_i = \sum_{i=1}^{2r+3} |y \cap (x \setminus v_i)| = |y \cap \bigcup_{i=1}^{2r+3} (x \setminus v_i) | \geq (2r+3)s - d_J(x,y)~.
$$

For each $i$, $v_i = x \setminus (x \setminus v_i) \cup (v_i \cap ([n] \setminus x))$,
where $x \setminus v_i \subseteq x$ and $(v_i \cap ([n] \setminus x)) \cap x = \varnothing$,
and hence by Lemma~\ref{lem:simple_dist} we have
$$
d_J (v_i,y) = |v_i \setminus y|=  |x \setminus y| +|(v_i \cap ([n] \setminus x)) \setminus y| -| (x \setminus v_i) \setminus y|
$$
which is equal to
$$
= d_J(x,y)+s-b_i - (s- a_i) = d_J(x,y) +a_i- b_i~.
$$
Therefore,
\begin{align*}
&\sum_{i=1}^{2r+3} d_J (y,v_i) = d_J (x,y) (2r+3) + \sum_{i=1}^{2r+3} a_i - \sum_{i=1}^{2r+3} b_i \geq (d_J(x,y) +s)(2r+3) - 2d_J(x,y) \\
& = (2r+1)d_J (x,y) +s(2r+3) \geq (2r+1)(r+1)+s(2r+3)  > (r+s) (2r+3),
\end{align*}
and hence there exists some $v_i \in C$ such that $d_J (y,v_i) > r+s$.

Thus, $y\not\in B(v_i,r+s)$ and the equality $B(x,r) = \bigcap_{i=1}^{2r+3} B(v_i,r+s)$ follows.

\end{IEEEproof}

\begin{theorem}
\label{lem:min_ServersJ2}
If $L\geq s(2r+3)$ in the Johnson scheme $J(n,L)$, then $N_J(L,r,s) \geq 2r + 3$.
\end{theorem}
\begin{IEEEproof}
Assume to the contrary that $N_J(L,r,s) \leq 2r + 2$, i.e., there exists a PPRIC code
$C=\{v_1,v_2,\ldots,v_{2r+2}\} \subseteq \cW^x_s$ which implies
that $B(x,r) = \bigcap_{c \in C} B(c,r+s)$.

Let $u_i = v_i \setminus x$ and $t_i = x \setminus v_i$ for $1 \leq i \leq 2r+2$, i.e.,
$v_i = (x \setminus t_i) \cup u_i$. For each $1 \leq i \leq r+1$ let $a_i \in t_i$ be any arbitrary
element in $\supp (x)$ which is contained in $\supp (v_i)$.
For each $r+2 \leq i \leq 2r+2$ let $b_i \in u_i$ be an arbitrary element
in $\supp (v_i)$ which is not contained in $\supp(x)$.

Let $\sigma$ be a subset of $x$ with size $r+1$ which contains $\{ a_i ~:~ 1 \leq i \leq r+1 \}$
(note that some of the $a_i$'s might be the same).
Similarly, let $\beta$ be a subset of $[n] \setminus x$ with size $r+1$
which contains $\{ b_i ~:~ r+2 \leq i \leq 2r+2 \} \cup \beta'$.

Define the word $y = x \setminus \sigma \cup \beta$.
Since $x \in J(n,L)$, $|\sigma|=|\beta|=r+1$, $\sigma \subset x$, and $\beta \cap x = \varnothing$, it follows that
$y \in J(n,L)$ and $d_J(x,y)=r+1$ and hence $y \notin B(x,r)$. Next, we compute a lower bound on the distance
between $y$ and each codeword of $C$.
Since $v_i = (x \setminus t_i) \cup u_i$, where $t_i \subseteq x$ and $u_i \cap x = \varnothing$,
it follows that Lemma~\ref{lem:simple_dist} can be applied to obtain
$$
d_J (v_i,y) = |x \setminus y| + |u_i \setminus y| - |t_i \setminus y|~.
$$
We distinguish between two cases:

\noindent
{\bf Case 1:} If $1 \leq i \leq r+1$, then
$|x \setminus y|=d_J(x,y)=r+1$, $|u_i|=s$ and hence $|u_i \setminus y| \leq s$. Furthermore, since
$a_i \in t_i$ and $a_i \notin y$, it follows that $|t_i \setminus y| \geq 1$ and therefore,
$$
d_J (v_i,y) = |x \setminus y| + |u_i \setminus y| - |t_i \setminus y| \leq r+1 +s-1 = r+s.
$$

\noindent
{\bf Case 2:} If $r+2 \leq i \leq 2r+2$, then
$|x \setminus y|=d_J(x,y)=r+1$, $|u_i|=s$ and since $b_i \in u_i \cap y$, it follows that $|u_i \setminus y| \leq s-1$. Furthermore,
$|t_i \setminus y| \geq 0$ and therefore,
$$
d_J (v_i,y) = |x \setminus y| + |u_i \setminus y| - |t_i \setminus y| \leq r+1 +s-1 = r+s.
$$

Thus, by Cases 1 and 2 we have that for each $1 \leq i \leq 2r+2$, $d(v_i,y) \leq r+s$, i.e.
$y \in \bigcap_{c \in C} B(c,r+s)$, a contradiction to $B(x,r) = \bigcap_{c \in C} B(c,r+s)$ since $y \notin B(x,r)$.
Thus, $N_J(L,r,s) \geq 2r + 3$.
\end{IEEEproof}

\begin{corollary}
\label{cor:min_ServersJ}
If $L\geq s(2r+3)$ in the Johnson scheme $J(n,L)$, then $N_J(L,r,s) = 2r + 3$.
\end{corollary}

PPRIC codes have their own combinatorial interest and not just as the minimum number of servers
in the PPR schemes. For the Johnson schemes, an $(n,L,k,t)$ \emph{covering code} $C$ is a collection
of binary words of length $n$ and weight $L$ with $L-k$ \emph{ones} in the first $L$ coordinates,
such that for each vector $v$ of length $n$ and weight $L$ with $L-t$ \emph{ones} in the first coordinates,
there exists exactly one codeword $c \in C$ such that $d_J(c,v)=k-t$. These types of codes (designs) were considered
for various problems, e.g.~\cite{Etz96}. Bounds on their size were not considered before, but they
can be easily derived from $(n,k,t)$ covering design.

A \emph{Steiner system} $S(t,k,n)$ is a collection of $k$-subsets (called \emph{blocks}) of an $n$-set,
such that each $t$-subset of the $n$-set is contained in exactly one block of the collection.
Clearly, a Steiner system $S(t,k,n)$ is a covering design $C(t,k,n)$.

One can easily verify the following result.
\begin{theorem}
\label{thm:direct_prod}
If $C_1$ is an $(L,k,t)$ covering design and $C_2$ is an $(n-L,k,t)$ covering design,
then $C \triangleq \bar{C}_1 \times C_2$ is an $(n,L,k,t)$ covering code $C$ in the Johnson scheme.
Moreover, if $C_1$ is a Steiner system $S(t,k,L)$ and $C_2$ is a Steiner system $S(t,k,n-L)$, then
$C$ is a covering code of minimum size.
\end{theorem}

Theorem~\ref{thm:direct_prod} implies a simple way to construct optimal $(n,L,k,t)$ covering codes.
But, generally a product construction for two covering designs in the Hamming scheme does not yield
an optimal covering code in the Johnson scheme. The simplest example for this claim can be viewed from
the minimum covering design of size $r+3$ implied by Corollary~\ref{cor:exactr+3} for the Hamming scheme,
and the minimum covering code of size $2r+3$ implied by Corollary~\ref{cor:min_ServersJ} for the Johnson scheme.
It is clear that this leaves lot of interesting questions concerning PPRIC codes, MIPPR words,
and covering codes in the Johnson scheme. These questions are left for future research.
The Grassmann scheme is even more difficult to analyse and it is left for future research as well.

\section{Conclusion and Future Research}
\label{sec:conclude}

This paper studies a new family of protocols, called private proximity retrieval,
which provide a certain form of private computation. Under this paradigm, the user has some file and
is interested to retrieve the indices of all files in the database which are close to his file.
As opposed to many existing privacy protocols, here perfect privacy cannot be achieved and the main goal
of the paper is to determine the tradeoff between the user's privacy and the number of servers in the system.
The construction of these protocols is mostly based on a new family of codes, called PPRIC codes, which have their own interest as covering codes.
In the binary Hamming scheme they are constant weight codes whose balls with radius $\ell$ intersect in exactly a ball of radius $r$.
These covering codes form a generalization for covering designs. Our exposition raises many problems for
future research, some of which were mentioned alongside our discussion. These problems are of interest
from both coding theory and combinatorics points of view.

\section*{Acknowledgement}

The authors are deeply indebted to Oliver Gnilke and David Karpuk for many helpful discussions and for their collaboration and contribution to the results of this work which were presented in~\cite{EGKYZ}.


\begin{thebibliography}{99}

\bibitem{BU18A}
K. Banawan and S. Ulukus, ``The capacity of private information retrieval from coded databases,'' in \emph{IEEE Transactions on Information Theory}, vol. 64, no. 3, pp. 1945-1956, March 2018.

\bibitem{BU18B}
K. Banawan and S. Ulukus, ``The capacity of private information retrieval from byzantine and colluding databases,'' in \emph{IEEE Transactions on Information Theory}.
doi: 10.1109/TIT.2018.2869154


\bibitem{BIK05}
A. Beimel, Y. Ishai, and E. Kushilevitz, ``General constructions for information-theoretic private information retrieval," \emph{Journal of Computer and System Sciences}, vol. 71, no. 2, pp.\,213--247, 2005.

\bibitem{BIKR02}
A. Beimel, Y. Ishai, E. Kushilevitz, and J.F. Raymond, ``Breaking the $O(n^{1/(2k-1)})$ barrier for information theoretic private information retrieval,"
\emph{Proc. of the 43rd Symposium on Foundations of Computer Science},  Vancouver, B.C., IEEE Computer Society, pp.\,261--270, 2002.

\bibitem{BIM00}
A. Beimel, Y. Ishai, and T . Malkin, ``Reducing the servers computation in private information retrieval: PIR with preprocessing," \emph{Proc. of the 20th Annual International Cryptology Conference LNCS 1880}, Santa Barbara, CA, Springer, pp.\,55--73, 2000.

\bibitem{CHY14}
T. H. Chan, S. Ho, and H. Yamamoto, ``Private information retrieval for coded storage," arXiv preprint arXiv:1410.5489 (2014).

\bibitem{CHY15}
T. H. Chan, S.W. Ho, and H. Yamamoto, ``Private information retrieval for coded storage," \emph{Proc. IEEE Int. Symp. on Inf. Theory}, pp.\,2842--2846, Hong Kong, Jun. 2015.

\bibitem{CFN08}
E. Chavez, K. Figueroa, and G. Navarro, ``Effective proximity retrieval by ordering permutations,"
\emph{IEEE Transactions on Pattern Analysis and Machine Intelligence} (\emph{TPAMI}), vol. 30, no. 9, pp. 1647--1658, 2008.


\bibitem{CKGS98}
{B. Chor, E. Kushilevitz, O. Goldreich, and M. Sudan},
``Private information retrieval," \emph{J.  ACM}, 45, 1998. Earlier version in FOCS 95.

\bibitem{DG14}
{Z.\,Dvir, and S.\,Gopi},
``2-server pir with sub-polynomial communication,"
arXiv:1407.6692v1, Jul. 2014.

\bibitem{Etz96}
     {T. Etzion,}
     ``On the nonexistence of perfect codes in the Johnson scheme'',
     {\em SIAM J. Discrete Math.,} 9, pp. 201--209, May 1996.

\bibitem{EGKYZ}
   {T. Etzion, O. Gnilke, D. Karpuk, E. Yaakobi, and Y. Zhang,}
     ``Private proximity retrieval'',
     \emph{2019 IEEE International Symposium on Information Theory (ISIT)}, Paris, France, 2019, pp. 2119--2123.

\bibitem{EtVa11}
     {T. Etzion and A. Vardy,}
     ``Error-correcting codes in projective space'',
     {\em IEEE Trans. on Inform. Theory,} 57, pp. 1165--1173, 2011.

\bibitem{EWZ95}
    {T. Etzion, V. Wei, and Z. Zhang,}
    ``Bounds on the sizes of constant weight covering codes'',
    {\em Design, Codes, and Cryptography}, 5, pp. 217--239, 1995.

\bibitem{FGHK17}
      R. Freij-Hollanti, O.W. Gnilke, C.Hollanti, and D.A. Karpuk, ``Private information
      retrieval from coded databases with colluding servers," arXiv:1611.02062v3, Aug. 2017.


\bibitem{Gor95}
    {D. M. Gordon,} https://www.dmgordon.org/cover.

\bibitem{GPK95}
    {D. M. Gordon, O. Patashnik, and G. Kuperberg,}
    ``New constructions for covering designs'',
    {\em Journal of Combinatorial Designs}, 3, pp. 269--284, 1995.


\bibitem{Dave18}
D. Karpuk, ``Private computation of systematically encoded data with colluding servers,'' \emph{2018 IEEE International Symposium on Information Theory (ISIT)}, Vail, CO, 2018, pp. 2112--2116.

\bibitem{KoKs08}
    {R. K\"{o}tter and F. R. Kschischang,}
    ``Coding for errors and erasures in random network coding'',
    {\em IEEE Trans. on Inform. Theory}, 54, pp. 3579--3591, 2008.

\bibitem{KRA16}
S. Kumar, H.-Y. Lin, E. Rosnes, A. Graell i Amat, ``Achieving maximum distance separable private information retrieval capacity with linear codes," arXiv:1712.03898v3, Aug. 2018.


\bibitem{MWS}
    {F. J. MacWilliams and N. J. A. Sloane,}
     \emph{The Theory of Error-Correcting Codes},
     Amsterdam: North-Holland Publishing Company, May 1978.

\bibitem{MY16}
Y.A. Malkov and D.A. Yashunin, ``Efficient and robust approximate nearest neighbor search using Hierarchical Navigable Small World graphs," arXiv:1603.09320, Mar. 2016.

\bibitem{Mil79}
    {W. H. Mills,}
    ``Covering designs I: coverings by a small number of subsets'',
    {\em Ars Combin.}, 8, pp. 199--315, 1979.

\bibitem{MiMu92}
    {W. H. Mills and R. C. Mullin,}
    ``Coverings and Packings'',
    {\em Contemporary design theory: a collection of surveys,}
    J. H. Dinitz and D. R. Stinson (Editors), Wiley, New York,
    pp. 371--399, 1992.

\bibitem{MAMA17}
    M. Mirmohseni and M.A. Maddah-Ali, ``Private function retrieval,'' arXiv:1711.04677v2, Nov. 2017.

\bibitem{MoVR}
    {M. Morley and G. H. van Rees,}
    ``Lottery schemes and covers'',
    {\em Utilitas Math.}, 79, pp. 159--165, 1990.

\bibitem{NTLHB11}
 A. Narayanan, N. Thiagarajan, M. Lakhani, M. Hamburg, and D. Boneh, ``Location privacy via private proximity testing," \emph{Proc. Network and Distributed System Security Symp.}, 2011.

\bibitem{NPS12}
J.D. Nielsen, J.I. Pagter and M.B. Stausholm, ``Location privacy via actively secure private proximity testing,"  \emph{IEEE Int. Conf. on Pervasive Computing and Comm. Workshops}, pp. 381--386, Lugano, 2012.


\bibitem{OLRK18}
S.A. Obead, H.-Y. Lin, E. Rosnes and J. Kliewer, ``Capacity of private linear computation for coded databases,'' arXiv:1810.04230, Oct. 2018.

\bibitem{PKB15}
C. Patsakis, P. Kotzanikolaou, and M. Bouroche, ``Private proximity testing on steroids: An NTRU-based protocol,"
\emph{Security and Trust Management}, pp. 172--184, 2015.

\bibitem{RK18}
N. Raviv and D. Karpuk, ``Private polynomial computation from lagrange encoding,'' arXiv:1812.04142, Dec. 2018.



\bibitem{Sch64}
J. Sch\"{o}nheim, ``On coverings,'' in \emph{Pacific J. math.}, vol. 14, pp. 1405 -- 1411, 1964.

\bibitem{SJ17A}
H. Sun and S. A. Jafar, ``The capacity of private information retrieval,'' in \emph{IEEE Transactions on Information Theory}, vol. 63, no. 7, pp. 4075--4088, July 2017.

\bibitem{SJ18A}
H. Sun and S. A. Jafar, ``The capacity of robust private information retrieval with colluding databases,'' in \emph{IEEE Transactions on Information Theory}, vol. 64, no. 4, pp. 2361-2370, April 2018.



\bibitem{SJ17B}
H. Sun and S.A. Jafar, ``The capacity of private computation,'' arXiv:1710.11098v3, Nov. 2017.



\bibitem{TGR18}
R. Tajeddine, O. W. Gnilke and S. El Rouayheb, ``Private information retrieval from {MDS} coded data in distributed storage systems,'' in \emph{IEEE Transactions on Information Theory}, vol. 64, no. 11, pp. 7081-7093, Nov. 2018.

\bibitem{Tod85}
    {D. T. Todorov,}
    ``On some covering designs'',
    {\em J. Combinatorial Theory, Ser. A}, 39, pp. 83--101, 1985.

\bibitem{WS16}
Q. Wang and M. Skoglund, ``Linear symmetric private information retrieval for MDS coded distributed storage with colluding servers," arXiv:1708.05673, Aug. 2017.

	

\bibitem{G04}
G. William,
``A survey on private information retrieval," Bulletin of the EATCS. 2004.

\bibitem{WY05}
{D.P. Woodruff and S. Yekhanin},
``A geometric approach to information-theoretic private information retrieval,"
\emph{IEEE Conference on Computational Complexity}, p.\,275--284, 2005.


\bibitem{Y10}
{S. Yekhanin},
``Private information retrieval,"
\emph{Comm. of the ACM}, vol.\,53, no.\,4, pp.\,68--73, 2010.

\bibitem{Y08}
{S. Yekhanin},
``Towards 3-query locally decodable codes of subexponential length,"
\emph{Joural ACM}, vol.\,55, no.\,1, pp.\,1--16, 2008.



\end{thebibliography}
\end{document}